\documentclass[10pt,conference]{IEEEtran}
\IEEEoverridecommandlockouts

\usepackage[tikz]{bclogo}
\usepackage{transparent}
\usepackage{tcolorbox}
\usepackage{dirtytalk}
\usepackage{comment}
\usepackage{booktabs}
\usepackage{array}
\usepackage{cite}
\usepackage{amsmath,amssymb,amsfonts}
\usepackage{graphicx}
\usepackage{adjustbox}
\usepackage{textcomp}
\usepackage{xcolor}
\usepackage{float}
\usepackage{xcolor,colortbl}
\usepackage{multirow}
\usepackage{subcaption} 
\usepackage{tcolorbox}
\usepackage{csquotes}
\usepackage[hidelinks]{hyperref}
\usepackage{threeparttable}
\usepackage{standalone}
\usepackage{balance}
\usepackage{threeparttable}
\usepackage{pbox}

\usepackage[linesnumbered,ruled,vlined]{algorithm2e}

\usepackage{bm}
\usepackage[utf8]{inputenc}
\usepackage{pgfplots}
\DeclareUnicodeCharacter{2212}{−}
\usepgfplotslibrary{groupplots,dateplot}
\usetikzlibrary{patterns,shapes.arrows}
\pgfplotsset{compat=newest}

\newcommand{\keystate}[1]{
\begin{tcolorbox}[leftrule=1mm,toprule=0mm,bottomrule=0mm,left=1pt,right=2pt,top=2pt,bottom=2pt]
\em #1
\end{tcolorbox}
}

\newcommand{\revisioncolor}{black} 
\newcommand{\revision}[1]{{\color{\revisioncolor}#1}}

\newenvironment{KEY}[1]%
{\noindent\begin{minipage}[c]{\linewidth}%
\begin{bclogo}[couleur=gray!30,%
                arrondi=0.1,%
                logo=\bclampe,%
                ombre=true]{\normalsize Key Characteristic} {#1}}%
{\end{bclogo}\end{minipage}\vspace{2mm}}

\def\BibTeX{{\rm B\kern-.05em{\sc i\kern-.025em b}\kern-.08em
		T\kern-.1667em\lower.7ex\hbox{E}\kern-.125emX}}

\usepgfplotslibrary{fillbetween}

\begin{document}


\title{Distilled Lifelong Self-Adaptation for Configurable Systems}	

\author{
\IEEEauthorblockN{Yulong Ye$^{1,2}$, Tao Chen$^{1,2 \ast}$, Miqing Li$^2$}

\IEEEauthorblockA{$^1$ IDEAS Lab, University of Birmingham, United Kingdom}
\IEEEauthorblockA{$^2$ School of Computer Science, University of Birmingham, United Kingdom}

\IEEEauthorblockA{yxy382@student.bham.ac.uk, t.chen@bham.ac.uk, m.li.8@bham.ac.uk}

\thanks{$^{\ast}$Corresponding author.}
}
	
	\maketitle

	\newcommand{\approach}{\texttt{DLiSA}}
 
	\begin{abstract}
		Modern configurable systems provide tremendous opportunities for engineering future intelligent software systems.
A key difficulty thereof is how to effectively \textcolor{black}{self-}adapt the configuration of a running system such that its performance (e.g., runtime and throughput) can be optimized under time-varying workloads. This unfortunately remains unaddressed in existing approaches as they either overlook the available past knowledge or rely on static exploitation of past knowledge without reasoning the usefulness of information when planning for self-adaptation. In this paper, we tackle this challenging problem by proposing \approach, a framework that self-adapts configurable systems. \approach~comes with two properties: firstly, it supports lifelong planning, and thereby the planning process runs continuously throughout the lifetime of the system, allowing dynamic exploitation of the accumulated knowledge for rapid adaptation. Secondly, the planning for a newly emerged workload is boosted via distilled knowledge seeding, in which the knowledge is dynamically purified such that only useful past configurations are seeded when necessary, mitigating misleading information. 




Extensive experiments suggest that the proposed \approach~significantly outperforms state-of-the-art approaches, demonstrating a performance improvement of up to 229\% and a resource acceleration of up to 2.22$\times$ on generating promising adaptation configurations. All data and sources can be found at our repository: \textcolor{blue}{\texttt{\url{https://github.com/ideas-labo/dlisa}}}.
	\end{abstract}
	
	\begin{IEEEkeywords}
		Self-adaptive systems, search-based software engineering, dynamic optimization, configuration tuning
	\end{IEEEkeywords}


\section{Introduction}
\label{section:introduction}

Software systems are often highly configurable~\cite{xu2015hey,DBLP:conf/sigsoft/0001L21,DBLP:journals/tosem/ChenL23a,DBLP:journals/pacmse/0001L24,chen2021mmo}. However, their operation environment is often confronted with dynamic and uncertain conditions that change over time~\cite{weyns2023towards,garcia2016towards}, which is crucial to their performance (e.g., runtime \cite{lesoil2021interplay}). Taking the \textsc{h2} database system as an example, its real-time workloads are known to highly fluctuate, prompting the system to dynamically adjust its configuration options to accommodate such changes \cite{muhlbauer2023analyzing}. 



\revision{To mitigate this, one promising way is to engineer self-adaptive configurable systems---a specific type of self-adaptive systems that, when the workload changes, self-adapt their configurations to meet different performance needs~\cite{salehie2009self,chen2022lifelong,DBLP:conf/seams/Chen22,jamshidi2017transfer2}.}  The critical challenge of self-adaptation lies in planning \cite{chen2020synergizing,chen2018adapt,chen2020search}, i.e., how to identify the most effective configuration (a.k.a. adaptation plan) amidst constantly changing workload at runtime. Recently, Search-Based Software Engineering (SBSE) has been considered a promising direction for solving this challenge, which tries to iteratively search for and refine configurations to locate the optimal one by using tailored search algorithms \cite{elkhodary2010fusion,gerasimou2018synthesis,ramirez2009applying,kinneer2021information}. The inherent search and optimization properties of SBSE make it well-suited to addressing the complexities of huge configuration spaces encountered in adapting configurable systems. Importantly, SBSE exhibits highly extensible potentials for complementing other approaches, such as control theoretical \cite{filieri2015automated,maggio2017automated,shevtsov2016keep} and learning-based methods \cite{chen2013self,ha2019deepperf,chen2014online, jamshidi2017transfer,chen2016self,zhang2023towards}, to achieve integrated schemes, thereby providing a comprehensive solution for runtime planning.

Beyond the exponentially growing search space and the non-linear interaction among configuration options, the ever-changing workload further intensifies the planning difficulties when self-adapting configurable systems. Particularly, the landscape of the search space may shift dramatically across different workloads, suggesting that a configuration optimized for one workload may become suboptimal or even perform poorly in another \cite{chen2022lifelong}. This dynamic nature requires planning to not only search for an optimal configuration under a newly emerged workload but should be doing so rapidly.

A promising resolution to that end is to reuse ``past knowledge'', i.e., configurations that were optimized under the previous workloads, for the planning to start working with under the current workload~\cite{kinneer2021information}. However, unfortunately, existing works often assume a stationary adaptation, which restarts the planning process from scratch following each workload change or at a fixed frequency. Such methods may be inefficient, as they fail to fully utilize historical search experiences, resulting in repetitive effort and a waste of valuable information that could inform more effective adaptation planning \cite{chen2018femosaa,elkhodary2010fusion,gerasimou2018synthesis}. \revision{Indeed, certain approaches have followed a dynamic adaptation~\cite{ramirez2009applying,chen2018effects,kinneer2021information} that exploit configurations found previously to speedup the planning (i.e., seeding). This, while running continuously, can still generate negative outcomes as the way how knowledge is exploited follows a static strategy: all (or randomly selected) configurations from the most recent past workload are used while any of those from earlier workloads are discarded. That said, the idea seems intuitive---the latest workload that has been changed may provide more useful information for the current newly emerged one while those older workloads may often be less useful. Yet, since the order of workload arriving in the system is uncertain, there is no guarantee that the configurations from the latest workload are all promising for the current one nor those from earlier workloads are completely irrelevant, as what has been implied in prior work~\cite{muhlbauer2023analyzing} and observed from our study in Section-II.C.}

To fill the above gap, in this paper, we propose a framework, dubbed {\approach}, to self-adapt configurable systems at runtime based on the MAPE-K loop \cite{kephart2003vision}. \revision{\approach~comes with a combination of two properties that makes it distinctive: (1) \textbf{lifelong planning}, where we leverage an evolutionary algorithm in the planning that runs continuously throughout the lifetime of the system while providing the foundation for seeding; and (2) \textbf{distilled knowledge seeding}---a truly dynamic knowledge exploitation strategy such that it not only seeds past configurations when there is evidence that they can be beneficial but also extracts the most useful ones to seed from all historical workloads, hence mitigating the misleading noises while keeping the most useful information. In this way, \approach~ensures that the exploitation of past knowledge is neither static nor completely abandoned, which fits with the characteristics of configurable systems.}


In a nutshell, our main contributions are as follows:
\begin{itemize}

       \item We show, by examples of configurable systems' landscapes, the key characteristic faced by self-adapting configuration at runtime under changing workloads.
 
	\item We develop a ranked workload similarity analysis to excavate correlations and patterns of past workloads, helping to extrapolate the traits of the new workload for more informed adaptation planning.
	
	\item We propose a weighted configuration seeding that distills the past knowledge, seeding only the most useful configurations and mitigating the misleading ones.
 
	
	\item \approach~is experimentally evaluated against four state-of-the-art approaches on nine real-world systems with different performance objectives, scales, and complexity, including 6--13 time-varying workloads. This leads to a total of 93 cases to investigate.
 
\end{itemize}

Experimental results encouragingly demonstrate that \approach~exhibits significant improvements in both efficacy (up to 2.29$\times$) and efficiency (up to 2.22$\times$).

The rest of this paper is organized as follows. Section \ref{section:background_and_motivation} introduces the background and motivation. Section \ref{section:methodology} provides the details of our proposed \approach. Section \ref{section:experimental_setup} presents our experiment methodology, followed by the experimental results in Section \ref{section:experimental_studies}. Section~\ref {section:discussion} discusses the most noticeable aspects of \approach. Threats to validity, related work, and conclusion are presented in Sections \ref{section:threats_to_validity}, \ref{section:related-work}, and \ref{section:conclusion}, respectively.

	\section{Background and Motivation} 
\label{section:background_and_motivation}

In this section, we discuss the preliminaries and main motivation of this work.


\subsection{Self-Adaptive Configurable Systems}

\revision{In this work, we focus on self-adaptive configurable systems. According to a well-known taxonomy~\cite{salehie2009self}, the self-adaptive configurable systems differ from the other concepts as follows:}

\revision{
\begin{itemize}
    \item \textbf{Self-Adaptive Systems:} These systems adapt to changes by modifying their behaviors, which could be any system's states (including structure and parameters) at runtime~\cite{salehie2009self}.
    \item \textbf{Self-Reconfigurable Systems:} A special type of self-adaptive systems that primarily alter their structure/architecture (including parameters) to adapt~\cite{garlan2004rainbow}.
    \item \textbf{Self-Adaptive Configurable Systems:} Unlike others, these systems specifically adapt by adjusting configuration parameters~\cite{wang2022agilectrl}.
\end{itemize}
}

\revision{Clearly, the self-adaptive configurable system is a type of self-adaptive/self-reconfigurable systems that primarily adjust configuration parameters at runtime to optimize their performance~\cite{salehie2009self,garlan2004rainbow,wang2022agilectrl}, focusing at the intersection between \textit{self-optimized} and \textit{self-configured} systems~\cite{salehie2009self}, which have been frequently studied in prior work involving dynamic workloads~\cite{chen2022lifelong,DBLP:conf/seams/Chen22,jamshidi2017transfer2}.}

\subsection{Problem Formalization}
\label{subsection:self-adaptation_planning_problem}
  
\begin{figure}[t]
\centering
   \includegraphics[width=0.35\textwidth]{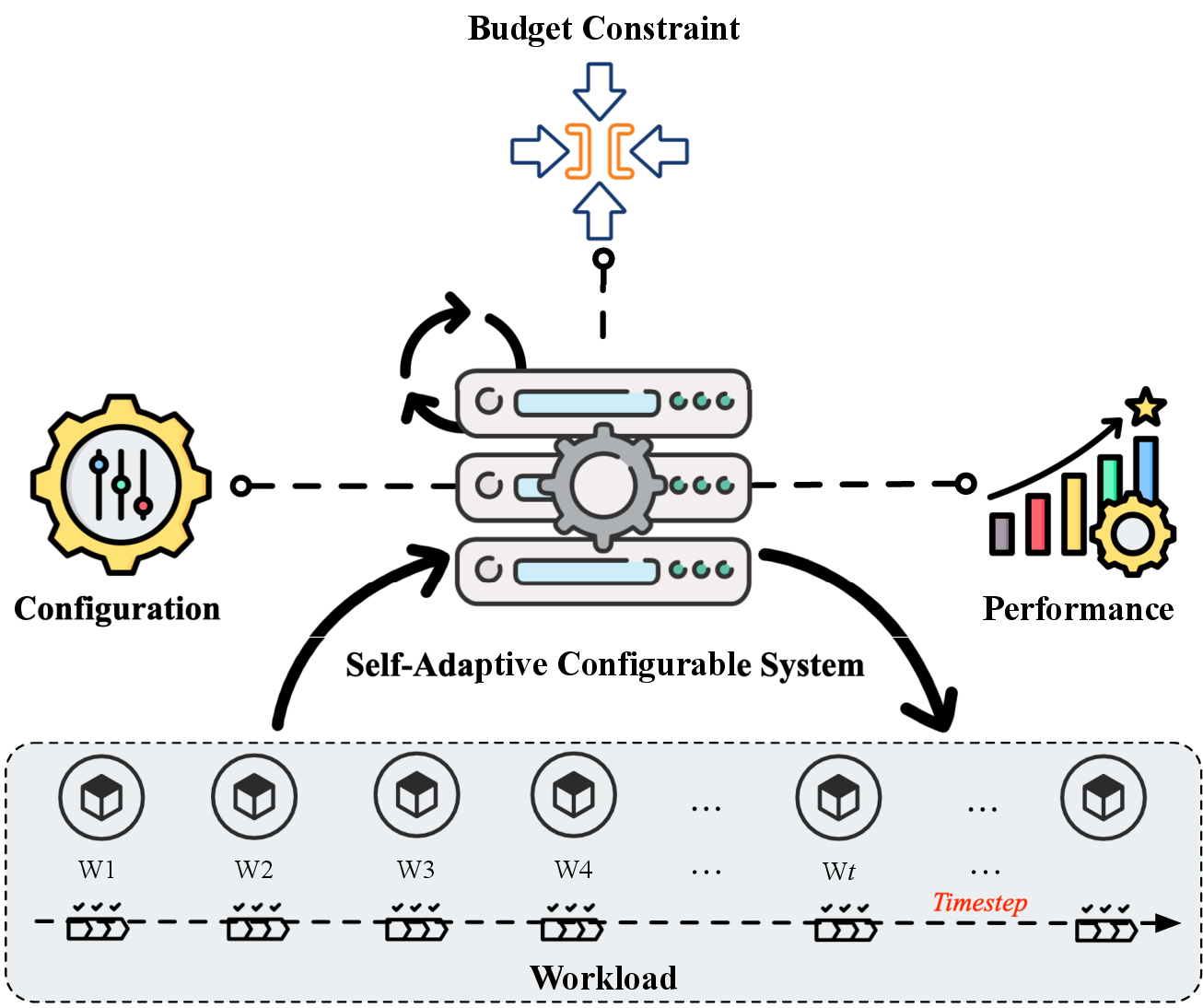}
    \caption{Self-adaptation planning for configurable systems.}
    \label{fig:problem_flowchart}
    \vspace{-3ex}
\end{figure}

Without loss of generality, self-adaptation planning for a given configurable system involves the following key concepts, as shown in Figure~\ref{fig:problem_flowchart}.

\begin{itemize}
    \item \textbf{System:} A configurable system with configurable options that can be adjusted at runtime.

    \item \textbf{Workload:} The time-varying and uncertain receiving jobs for which the system handles. The concrete instance can vary. For example, the workloads refer to different types and volumes of queries that emerged for database system \textsc{H2}; for file compressors such as \textsc{kanzi}, this becomes the incoming files to be compressed, which could be of diverse formats and sizes. 


    \item \textbf{Configuration:} An instance of variability for a configurable system, formed by a set of values for the configurable options. In this work, we consider both the configurations (and options) that require system rebooting and those that do not.
    
    
    \item \textbf{Performance:} \revision{The metric(s) that evaluates the behavior of the system, such as runtime (i.e., the time taken by the system to process a given workload) and throughput.} 
    

    \item \textbf{Budget constraint:} The budget of cost allowed for self-adaptation planning under the workload for a particular timestep. While the definition of budget varies, in this work, we use the number of system measurements during planning as the budget, \revision{which means that we can only measure a certain number of configurations as our constraint.} The measurement is chosen because: (1) it is independent of the implementation, such as language and hardware; (2) it eliminates the interference of clock time caused by the running system to be adapted, when it is run at the same machine as the planning process; (3) it has been widely used in existing work~\cite{DBLP:journals/tse/Nair0MSA20,DBLP:conf/sigsoft/0001L21}.
    
\end{itemize}

\revision{When a single performance objective is of concern,} the goal of planning when self-adapting a configurable system $S$ is, for each timestep $t$ in which the system handles a workload over the time horizon, \revision{to identify a configuration that optimizes the specific performance attribute, e.g., minimizing runtime or maximizing throughput,} of the target system, subject to a budget constraint for planning. \revision{Formally, this can be defined as:
\begin{equation}
	\begin{array}{l}
		\arg\min f_t(\textbf{\textbf{x}}) \text{ or } \arg\max f_t(\textbf{\textbf{x}}), \\
	\text{s.t. } r_t \leq R_t,
	\end{array}
\end{equation}}where \textbf{x} = ($x_1, x_2, \dots, x_n$) is a configuration with the values of $n$ options (e.g., $x_n$) in search space \bm{$\mathcal{X}$}. \revision{$f_t$ represents the performance attribute of the target system.} $r_t$ and $R_t$ respectively denote the cost consumption and the budget allowed for planning at timestep $t$.



\subsection{Motivation and Challenges}
\label{subsection:motivation_and_challenges}

\begin{figure}[t] 
	\centering
	\begin{subfigure}{0.23\textwidth} 
		\centering
		\includegraphics[width=4.5cm]{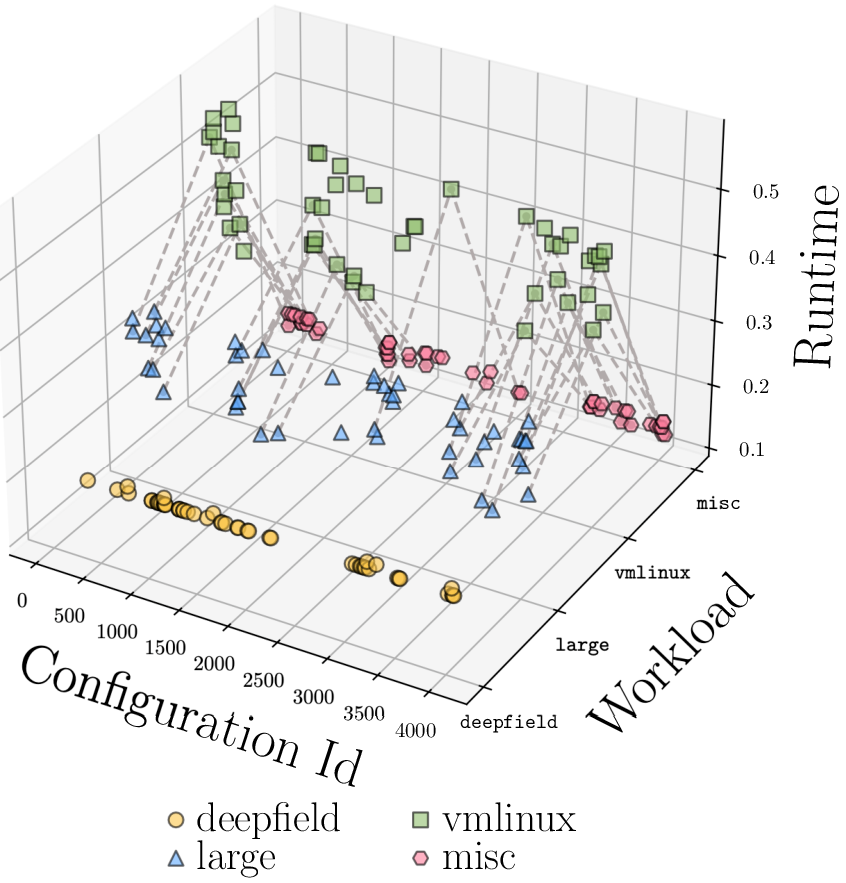} 
		\caption{\textsc{kanzi}}
		\label{fig:motivation_sub_a}
	\end{subfigure}
	\hspace{0.1cm}
	\begin{subfigure}{0.23\textwidth} 
		\centering
		\includegraphics[width=4.5cm]{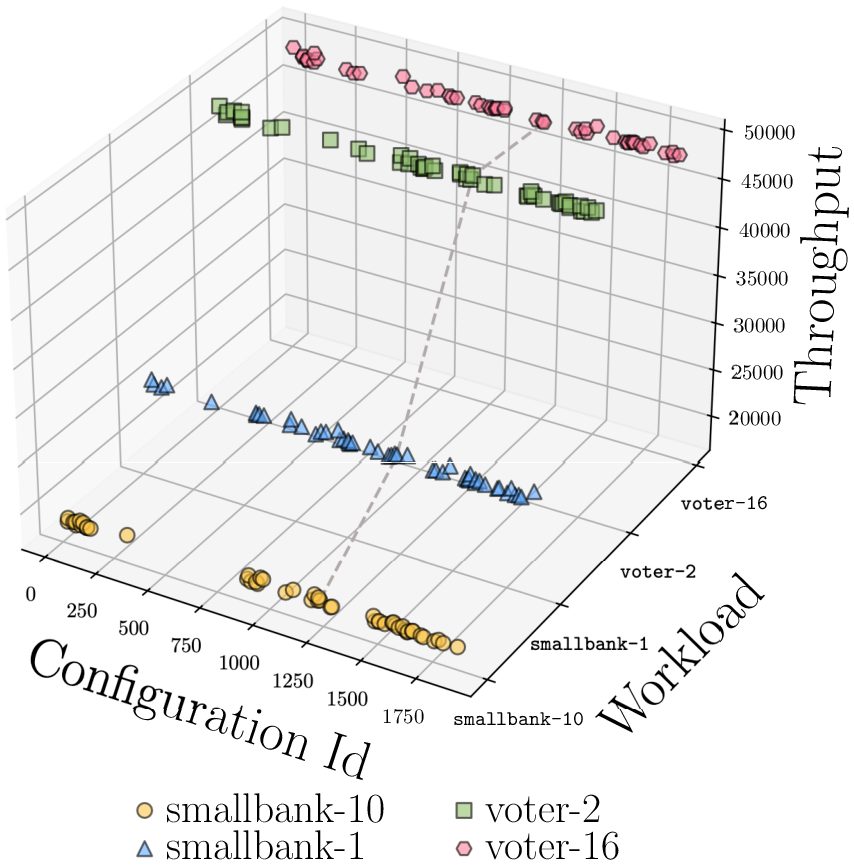} 
		\caption{\textsc{h2}}
		\label{fig:motivation_sub_b}
	\end{subfigure}
	\caption{An illustration of similarities and discrepancies in top 50 performing configurations across workloads. The same configurations are connected by dashed lines.}
        \label{fig:motivation}
        \vspace{-0.1cm}
\end{figure}

\revision{It is well-known that, when self-adapting configurable systems, the configurations produced under one workload might be useful to the other workloads~\cite{jamshidi2017transfer,DBLP:conf/seams/Chen22,muhlbauer2023analyzing}. However, it remains unclear how to explicitly extract the key knowledge and whether there exists irrelevant or even misleading information, i.e., noises. To uncover these underlying issues,} we analyze the datasets collected from commonly used configurable systems and their workload from prior studies \cite{alves2020sampling,velez2020configcrusher,weber2021white,muhlbauer2023analyzing}. The goal is to investigate what are the similarities and discrepancies between the configuration landscapes of different workloads. Figure~\ref{fig:motivation} shows the top 50 performing configurations for two systems under different workloads and we observe the following patterns (similar observations exist in other systems):

\begin{itemize}
    \item There could be a strong overlap of the promising configurations across workloads (the connected points). That said, a promising configuration under a workload could also be promising under the others. For example, the points connected by dashed lines for workloads \texttt{large}, \texttt{vmlinux}, and \texttt{misc} of \textsc{kanzi} in Figure~\ref{fig:motivation_sub_a}.  
    \item It is also possible that the promising configurations for each individual workload differ significantly. For instance, the points under workloads of \textsc{h2} in Figure~\ref{fig:motivation_sub_b} rarely overlap with each other. The same phenomenon occurs even for the workloads of the same system, e.g., workload \texttt{deepfield} against the others for \textsc{kanzi}. 
\end{itemize}

The above leads to a key characteristic for configurable systems, which motivates our work:

\begin{KEY}{
\textit{Top-performing configurations between workloads can be very similar or very discrepant, depending on both the systems (e.g., \textsc{kanzi} and \textsc{h2}) and the workloads within a single system (e.g., between workload \texttt{deepfield} and the others for \textsc{kanzi}).}}
\end{KEY}

Since the order of workloads arriving at a system is uncertain, the above suggests that ``seeding''\footnote{Seeding is a mechanism that benefits the planning for the current workload by reusing configurations optimized under the past workloads~\cite{chen2018effects,kinneer2021information}.} promising configurations optimized for the past workloads to the planning under the current workload can be beneficial, as long as we can:

\begin{itemize}
    \item \textbf{Challenge 1:} extract the most useful configurations discovered previously (the  configurations that are promising across workloads), if any, while doing so without injecting misleading information (the  configurations that are ``good'' under the past workloads only);
    \item \textbf{Challenge 2:} and detect when it is generally more harmful to seed than simply restart planning.
\end{itemize}

Nevertheless, existing approaches have failed to explicitly handle the above characteristics and challenges of configurable systems when running under changing workloads: on one hand, \textit{stationary adaptation} approaches (e.g., \texttt{FEMOSAA}~\cite{chen2018femosaa}) restarts a new search/planning from scratch with each workload change, but clearly, according to the above characteristic, this would waste the valuable knowledge from the past workload instances that could have been exploited~\cite{liu2023towards}. \revision{On the other hand, the \textit{dynamic adaptation} approaches (e.g., \texttt{Seed-EA}~\cite{kinneer2021information}) rely on a static assumption for the knowledge exploitation strategy: all configurations accumulated to the most recent past workload are useful for seeding, while those from previous ones are discarded. Besides, they always trigger seeding even when the benefits are unjustified. Because the seeds retain the planning state, indeed, the dynamic approaches can also be executed in a ``lifelong'' manner where the planning process runs continuously and adapts to the changes in the workload, but they may pick up potentially misleading information (optimal configurations found in the most recent past workload but are no longer promising) or missing useful hints (generally promising configurations found under earlier workloads), hindering the planning in the current workload.}

\begin{table}
\caption{\revision{Comparing \approach~against other approaches.}}
\label{tb:novelty}
\setlength{\tabcolsep}{0.8mm}
\begin{adjustbox}{width=\columnwidth,center}

\begin{tabular}{lllll}
\toprule
\textbf{Approach} & {\textbf{Knowledge Exploitation}} & {\textbf{Seeding}} & {\textbf{Workloads}} & {\textbf{Configurations}} \\ 
\midrule

\texttt{FEMOSAA}~\cite{chen2018femosaa}&N/A&N/A&N/A&N/A\\
\texttt{Seed-EA}~\cite{kinneer2021information}&Static&Always&Most recent past&All\\
\texttt{D-SOGA}~\cite{deb2007dynamic}&Static&Always&Most recent past&Random\\
\texttt{LiDOS}~\cite{chen2022lifelong}&Static&Always&Most recent past&All\\
\approach&Dynamic&On-demand&All historical&Distilled\\

\bottomrule
\end{tabular}
\end{adjustbox}
\end{table}




\revision{The above, therefore, are the key challenges and limitations we address in this paper. Table~\ref{tb:novelty} summarizes the novelty of \approach~against the properties of state-of-the-art approaches.}

\section{The \approach~Framework}
\label{section:methodology}

To tackle the current limitation and handle the key characteristic/challenges discussed in Section \ref{subsection:motivation_and_challenges}, we propose \approach---a distilled lifelong planning framework for self-adapting configurable systems with time-varying workloads. 

\approach~comes with two unique properties:

\begin{itemize}
    \item \textbf{Lifelong planning:} The planning runs continuously and adapts to workload changes---a typical case of dynamic optimization~\cite{DBLP:journals/swevo/NguyenYB12}---in which the state optimized across different workloads can be preserved. This provides the foundation for addressing \textbf{Challenge 1}.

    \item \textbf{Distilled knowledge seeding:} The knowledge of seeding is dynamically distilled, i.e., \approach~extracts the most useful configurations from all past workloads to seed into the current planning process; or triggers randomly-initialized planning from scratch when the overall distilled knowledge is deemed not sufficiently useful. This tackles both \textbf{Challenge 1} and \textbf{Challenge 2}.
\end{itemize}

Next, we will articulate \approach's designs in great detail.

\subsection{Architecture Overview}
\label{subsection:architecture_overview}

\begin{figure}[t]
    \centering
	\includegraphics[width=0.44\textwidth]{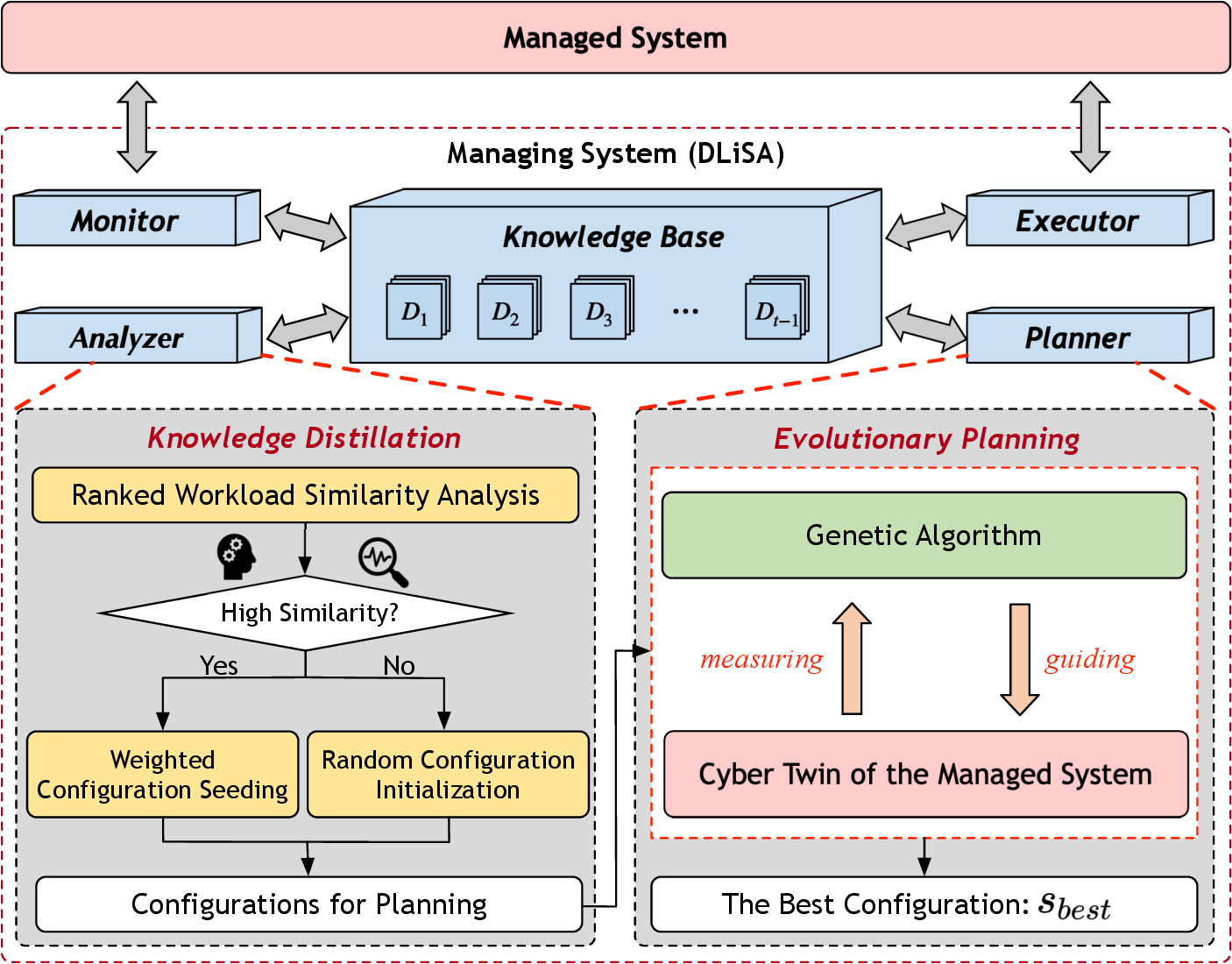}
	\caption{\approach~architecture for configurable systems.}
	\label{fig:algorithm_outline}
\end{figure}

We design \approach~using the typical MAPE-K architecture~\cite{kephart2003vision}, as shown in Figure~\ref{fig:algorithm_outline} and \textbf{Algorithm~\ref{alg:general_framework}}. In a nutshell, MAPE-K distinguishes two sub-systems---the managed system refers to the configurable systems that should be managed; and the managing system governs the self-adaptation, i.e., \approach. Once a workload change has been detected (e.g., a new incoming job), the \textit{Monitor} informs the \textit{Analyzer} to analyze the current and past status, which then triggers \textit{Planner} for reasoning about the best self-adaptation plan (configuration), subject to a given budget constraint. Finally, the best-optimized configuration is set to the managed system via \textit{Executor}. The \textit{Knowledge} refers to the preserved data that can be used by any phases in the MAPE loop. In this work, the knowledge we retain is the workloads experienced by the systems and all the corresponding configurations that were measured/discovered in the planning previously (line 7).

\approach~specializes two key phases in MAPE-K (lines 5-6):

\begin{itemize}
	\item \textbf{Analyzer:} The \textit{Knowledge Distillation} component identifies whether seeding is beneficial, and if that is the case, extracts the most useful configurations preserved previously to seed the planning under the current workload, realizing \textbf{distilled knowledge seeding} (see Section \ref{subsection:knowledge_distillation}).
	\item \textbf{Planner:} Here, we leverage \textit{Evolutionary Planning} component to evolve the configurations into better ones in the search space based on the given seeds, if any. We adopt a population-based optimizer where a set of the most promising configurations found is preserved and the best one is used for self-adaptation under a workload (see Section \ref{subsection:evolutionary_planning}). In particular, the search-based planning is conducted on a sandbox, which is often a Cyber-Twin or a surrogate system model, that allows expedited measurement of the configurations while emulating the behavior of the managed system under the given workload~\cite{DBLP:conf/ease/AhlgrenBDDGGHLL21,DBLP:journals/jid/WeynsBVYB22}. Through those seeds, the planning status for the previous workloads can be kept, hence rendering the entire process as \textbf{lifelong planning} over the time horizon. 
\end{itemize}




\SetKwInput{KwDeclare}{Declare}
\definecolor{note_color}{HTML}{FF4F78}
\begin{algorithm}[t]
	\caption{\approach~Framework}
	\label{alg:general_framework}
	\SetAlgoLined
	\footnotesize
        \KwIn{Cyber-Twin of the managed system \bm{$\mathcal{S}$}; the budge constraint $R_t$; threshold for triggering seeding $\alpha$; population size $N$.}
        \KwDeclare{The best configuration at current workload $s_{best}$; the set of configurations preserved at workload $t$, \textbf{D}$_t$. Seeded set of configurations \textbf{P}, and the \textit{knowledge base} \bm{$\mathcal{K}$}.}
	\While{the managed system is running}{
		$t=0$\\
		\If{workload change}{
			$t$ = $t + 1$\\
   \textbf{P} = \textsc{knowledgeDistillation}(\bm{$\mathcal{K}$}, $\alpha$, $N$) \\ 
    \textbf{D}$_t$, $s_{best}$ $\gets$ \textsc{evolutionaryPlanning}(\bm{$\mathcal{S}$}, \bm{$\mathcal{K}$}, \textbf{P}, $R_t$) \\
       \bm{$\mathcal{K}$} = \bm{$\mathcal{K}$} $\cup$ \textbf{D}$_t$\\
		\textsc{sendForAdaptation}($s_{best}$)\\		
		}
	}
\end{algorithm}

\subsection{Knowledge Distillation}
\label{subsection:knowledge_distillation}
As shown in Figure~\ref{fig:algorithm_outline}, with the dynamic exploitation strategy realized by knowledge distillation, \approach~seeks to distill the configurations optimized for all past workloads in two steps during planning: firstly, it selects representative configurations evaluated to assess the overall similarity amongst their performance across the workloads using a ranked similarity metric at the \textit{workload level}. A high value of the metric represents a higher likelihood of the seeding being useful. Next, if there is convincing evidence that seeding can be beneficial for planning, we probabilistically extract $N$ most useful configurations amongst those preserved for seeding using a quality weight at the \textit{configuration level}; otherwise, a random initialization process is used instead. An algorithmic illustration has also been shown in \textbf{Algorithm~\ref{alg:knowledge_distillation}}.

\textbf{\textit{Ranked Workload Similarity Analysis (When to seed?):}} \revision{For the trigger of seeding, the idea is that, if the majority of those configurations that were discovered under the past workloads are ``similarly good'', then it is likely that there is a strong chance for certain promising configurations optimized previously to be equally good under the current, newly emerged workload, i.e., the seeding should be beneficial.} 

As a result, we propose a ranked workload similarity analysis at the workload level using all the common configurations searched across workloads (including those that were ruled out). In particular, we quantify the similarity level between two workloads by using a pairwise ranking loss \cite{li2021mfes}. The rationale is that while the concrete performance of a configuration may fluctuate with workload changes, the relative rankings can remain indicative of similarity while being scale-free.

\begin{algorithm}[t]
	\caption{\textsc{knowledgeDistillation}}
	\label{alg:knowledge_distillation}
	\SetAlgoLined
	\footnotesize
	\KwIn{The knowledge base \bm{$\mathcal{K}$}; threshold for triggering seeding $\alpha$; initial/population size $N$}
	
	\KwOut{The set of initial configuration for planning \textbf{P}}
  \tcc{\textcolor{blue}{Ranked workload similarity analysis }}
	\For{$t$ = \rm 1 to \textsc{size}(\bm{$\mathcal{K}$})$\mathbf{-}$1}{
		\textbf{D}$_{t}^{t+1}$ $\gets$ common configurations from those evaluated in the planning of two adjacent workloads \textbf{D}$_{t}$ and \textbf{D}$_{t+1}$ from \bm{$\mathcal{K}$}\\
		$S_{t}^{t+1} \gets$ calculate the similarity between workloads $t$ and $t+1$ by (\ref{eq:2}) and (\ref{eq:3})\\
		$S_{sum}$ = $S_{sum}$ + $S_{t}^{t+1}$
	}
	$S_{ave}$  = \textsc{averaging}($S_{sum}$)\\
	   \tcc{\textcolor{blue}{Weighted configuration seeding}}
	\eIf{$S_{ave}$ $\geq$ $ \alpha$ and \textsc{size}(\bm{$\mathcal{K}$})$>0$}{
 \textbf{C} $\gets$ the $N/2$ best configurations under each workload \\
		\ForEach{$\forall c \in$ \textbf{C}}{
    \tcc{\textcolor{blue}{Get quality weight for $c$ by (\ref{eq:4})-(\ref{eq:6})}}
			\textbf{C$_{w}$} $\gets$ $\langle c,w_{c} \rangle$ $\gets$ $w_{c}$ = $w_{c,r}$ + $w_{c,t}$
		}
		
		\textbf{P} $\gets$ stochastically pick $N$ configurations from \textbf{C$_{w}$} using $w_{c}$\\
            \tcc{\textcolor{blue}{Only happens with one past workload}}
            \If{$|$\textbf{P}$| \neq N$} {
               \textbf{P} $\gets$ randomly initialize configurations till $|$\textbf{P}$| = N$
            }
	}
	{
		\textbf{P} $\gets$ randomly initialize configurations till $|$\textbf{P}$| = N$
	}
	\Return \textbf{P}
\end{algorithm}

We do so via the following steps (lines 1-6):

\begin{enumerate}
	\item For every pair of adjacent workloads (e.g., $t$ and $t+1$), retrieve all the evaluated configurations \textbf{D}$_t$ and \textbf{D}$_{t+1}$.
	\item Identify their common configurations \textbf{D}$_{t}^{t+1}$ (line 2).
	\item Compute the ranking loss by quantifying the number of misranked pairs in \textbf{D}$_{t}^{t+1}$ (line 3):
	\begin{equation}
            \hspace{-0.5cm} 
		\label{eq:2}
		\footnotesize
		\mathcal{L}(\textbf{D}_{t}^{t+1})=\sum_{j=1}^{N_{t}^{t+1}} \sum_{k=1}^{N_{t}^{t+1}} \mathbf{1}((f_t(\textbf{x}_j)<f_t(\textbf{x}_k)) \oplus (f_{t+1}(\textbf{x}_j)<f_{t+1}(\textbf{x}_k))),
	\end{equation}
	whereby $\oplus$ is the exclusive-or operator; $N_{t}^{t+1}$ is the number of configurations evaluated in both workloads $t$ and $t+1$ (i.e., the size of \textbf{D}$_{t}^{t+1}$). \revision{The ranking loss $\mathcal{L}(\textbf{D}_{t}^{t+1})$ represents the number of misranked pairs of configurations between adjacent workloads at timesteps $t$ and $t+1$, reflecting the discrepancy among them.}
	
	\item Assess the similarity between $t$ and $t+1$ ($S_{t}^{t+1}$) using the percentage of the order-preserving pairs, as follows:
	\begin{equation}
		\label{eq:3}
		S_{t}^{t+1} = 1 - \frac{\mathcal{L}(\textbf{D}_{t}^{t+1})}{N_{pairs}},
	\end{equation}
	where $N_{pairs}$ is the number of configuration combination in \textbf{D}$_{t}^{t+1}$ (line 3). 
	\item Calculate the average similarity score for all pairs of adjacent workloads, i.e., $S_{sav}$ (line 6).

\end{enumerate}

The seeding is said to be beneficial and should be triggered only if $S_{sav} \geq \alpha$, where $\alpha$ is a given threshold. Note that, if no common configurations are found between a pair of adjacent workloads, we set their similarity $S_{t}^{t+1}$ with a random value that is less than $\alpha$, serving as a reasonable guess when no reliable information can be extracted.

\textbf{\textit{Weighted Configuration Seeding (What to seed?):}} When \approach~determines that the seeding is necessary, we need to further select the most useful configurations for seeding the current workload. \revision{As observed in Section \ref{subsection:motivation_and_challenges}, there could be a strong overlap of good configurations across different workloads, yet sometimes, the promising ones for different workloads can be highly discrepant. Our idea here is to design a weighting scheme, such that it can discriminate the past configurations based on the likelihood of them being promising under the current workload. To this end,} we design a two-stage weighted seeding that operates at the configuration level, considering only the good configurations preserved. As such, we say a past configuration is useful for seeding if (1) it is good within its own workload (line 8) while (2) being robust and timely across all past workloads (lines 9-12).

The first stage---the local stage weighting---focuses on selecting the best configurations locally (based on the performance objective) under each workload. This is because those configurations that perform badly in a workload would be less meaningful for seeding. To that end, we filter the preserved configurations at each workload by 50\%, i.e., only $N \over 2$ configurations are considered where $N$ is the number of configurations to be seeded in the end (line 8).

In the second stage, we seek to globally weight the configurations across all the past workloads. The hypotheses are:

\begin{itemize}
	\item preserved configurations that have demonstrated robustness in many past workloads (as they were not ruled out) are likely to perform well in the new workload;
	\item since planning under a later workload might have evolved by integrating previously accumulated knowledge, configurations preserved in such a later workload are likely to exhibit good performance in the new workload.
\end{itemize}

Therefore, we use quality weight to sort the previously selected configurations from the first stage and it has two components: a robustness weight and a timeliness weight (lines 9-12). Specifically, a robustness weight is allocated to each configuration based on its recurrence across multiple past workloads (line 10). Configurations that appear in a larger number of workloads receive higher robustness weights, reflecting their robustness for being preferred frequently and the likelihood of successful performance:

\begin{equation}
	\label{eq:4}
	w_{c,r} = \frac{O_{c}}{H},
\end{equation}
where $O_{c}$ is the count of past workloads in which the configuration $c$ is preserved and $H$ denotes the total number of past workloads. In contrast, the timeliness weight of a configuration is calculated based on the chronological occurrence of the latest workload where the configuration is preserved (line 10). Configurations associated with more recent workloads are presumed to have integrated prior knowledge and are thus given higher timeliness weights:
\begin{equation}
	\label{eq:5}
	w_{c,t} = \frac{S_{c}}{H},
\end{equation}
where $S_{c}$ is the sequential number of the latest workload that the configuration $c$ is associated with, indicating the most recent (largest) order in which the configuration appears across the past workloads.

Since both $w_{c,r}$ and $w_{c,t}$ range between 0 and 1, the quality weight of configuration $c$ is then computed as:
\begin{equation}
	\label{eq:6}
	w_c = w_{c,r} + w_{c,t}
\end{equation}

In the end, we stochastically select $N$ configurations according to $w_c$ for seeding, where a greater value of $w_c$ stands a higher probability of being favored. This, compared with selecting them deterministically, still retains a low possibility of selecting ``less useful'' configurations for seeding, hence maintaining diversity to escape from the local optima.

Notably, when there is exactly one previous workload, only the first stage would work as the number of configurations to be chosen ($N \over 2$)  is smaller than the number required ($N$). The remaining $N \over 2$ configurations are then randomly generated. Of course, there will be no seeding under the very first workload.

\subsection{Evolutionary Planning}
\label{subsection:evolutionary_planning}

As mentioned, \approach~works the best with using evolutionary algorithms for planning because (1) they are based on population which fits well with the seeding---it caters to a set of configurations instead of one; (2) they have been widely studied in SBSE/self-adaptation~\cite{DBLP:journals/csur/HarmanMZ12,DBLP:journals/jid/WeynsBVYB22}. In this work, we employ Genetic Algorithm (GA) \cite{back1993overview} for seeded planning. In a nutshell, GA works by iteratively reproducing from promising configurations via crossover and mutation, as evaluated on the Cyber-Twin (see Section \ref{subsection:component_and_parameter_settings}), to evolve into even better ones. We adopt elitist-based GA where only the top-performing configurations are preserved in each iteration. Since GA has also been used for tuning configurable systems, interested readers can refer to prior work for a more detailed elaboration \cite{ramirez2009applying,chen2022lifelong,DBLP:conf/ssbse/BowersFC18}.

	\section{Experimental Setup}
\label{section:experimental_setup}

We experimentally assess the performance of {\approach~}by unraveling the following research questions (RQs):

\begin{itemize}
	\item {\bf RQ1:} How effective is {\approach~}against state-of-the-art approaches? 
	\item {\bf RQ2:} How efficient is {\approach~}compared with others? 
	\item {\bf RQ3:} What benefits do ranked workload similarity analysis and weighted configuration seeding each provide?
	\item {\bf RQ4:} How does $\alpha$ affect {\approach}'s performance?
\end{itemize}

All experiments are run in a Python environment on MacOS with a quad-core 1.4 GHz CPU and 8GB RAM.

\subsection{Subject Systems, Workloads, and Configurations}

\subsubsection{Systems}\revision{We follow all the systems from a prior empirical study \cite{muhlbauer2023analyzing} as our subjects, which investigates a range of widely studied configurable systems \cite{alves2020sampling,velez2020configcrusher,weber2021white}. Our selection aligns with this study to ensure consistency and comprehensiveness.} Specifically, these systems are carefully chosen to span a variety of application domains, performance objectives, and programming languages, including both Java and C/C++, thereby providing a solid foundation for our investigation into systems with diverse characteristics, as shown in Table \ref{tb:subject_sas}. \revision{More details on how to use these systems for conducting experiments can be found in \cite{muhlbauer2023analyzing,workload_similarity_dataset}.}

\begin{table}
	\centering
	\footnotesize
	\tabcolsep=0.03 cm
	\caption{Subject system characteristics. The details of their workloads can be found in our repository: \begin{small}\revision{\texttt{\url{https://github.com/ideas-labo/dlisa}}}\end{small}.}
	\label{tb:subject_sas}
	\begin{threeparttable}
		\begin{tabular}{lllllccc}
\toprule
\textbf{System} & \textbf{Lang.} & \textbf{Domain} & \textbf{Perf.} & \revision{\textbf{Version}}             & \textbf{\#O} & \textbf{\#C} & \textbf{\#W} \\ \midrule
JUMP3R\revision{\cite{subjectsys_jump3r}}         & Java              & Audio Encoder   & Runtime              & \revision{1.0.4}                        & 16           & 4196         & 6            \\
KANZI\revision{\cite{subjectsys_kanzi}}           & Java              & File Compressor & Runtime              & \revision{1.9}                          & 24           & 4112         & 9            \\
DCONVERT\revision{\cite{subjectsys_dconvert}}        & Java              & Image Scaling   & Runtime              & \revision{1.0.0-$\alpha$7} & 18           & 6764         & 12           \\
H2\revision{\cite{subjectsys_h2}}              & Java              & Database        & Throughput           & \revision{1.4.200}                      & 16           & 1954         & 8            \\
BATLIK\revision{\cite{subjectsys_batlik}}          & Java              & SVG Rasterizer  & Runtime              & \revision{1.14}                         & 10           & 1919         & 11           \\ \midrule
XZ\revision{\cite{subjectsys_xz}}              & C/C++             & File Compressor & Runtime              & \revision{5.2.0}                        & 33           & 1999         & 13           \\
LRZIP\revision{\cite{subjectsys_lrzip}}           & C/C++             & File Compressor & Runtime              & \revision{0.651}                        & 11           & 190          & 13           \\
X264\revision{\cite{subjectsys_x264}}            & C/C++             & Video Encoder   & Runtime              & \revision{baee400…}                     & 25           & 3113         & 9            \\
Z3\revision{\cite{subjectsys_z3}}              & C/C++             & SMT Solver      & Runtime              & \revision{4.8.14}                       & 12           & 1011         & 12           \\ \bottomrule
\end{tabular}
		\scriptsize
		\#\textbf{O:} No. of options;
		\#\textbf{C:} No. of configurations;
		\#\textbf{W:} No. of workloads tested.
	\end{threeparttable}
\end{table}

\subsubsection{Workloads}\revision{The workloads we studied are diverse and domain-specific.} For example, for SMT solver \textsc{z3}, the workload can be different SMT instances that are of diverse complexity, e.g., \texttt{QF\_RDL\_orb08} and \texttt{QF\_UF\_PEQ018}; for database system \textsc{h2}, the workloads are requests with different rates and types (e.g., read-only or read-write), such as \texttt{tpcc-2} and \texttt{ycsb-2400}. \revision{In this study, we use the same various workloads as in \cite{muhlbauer2023analyzing}, which range from 6 to 13 depending on the systems (denoted as W1, W2,\dots, W13). Self-adaptations are triggered as those different workloads arrive at the system in a certain order. Here, we randomly shuffle the order of all workloads to arrive at a system and test self-adaptation therein.} 
\subsubsection{Configurations}\revision{Each system in our study features a distinct configuration space, covering different option types (e.g., integers, boolean, and enumerates) and dimensions.}

\revision{Overall, these diverse systems and workloads provide a robust foundation for assessing the efficacy and efficiency of our approach across different contexts and configurations.}





\subsection{Compared Adaptation Approaches}
\label{subsection:compared_adaptation_approaches}
For the comparative analysis in our study, we compare \approach~with the following state-of-the-art approaches:

\begin{itemize}
	\item \textbf{\texttt{FEMOSAA}\footnote{We use the single-objective version and pair it with GA.} (Stationary Adaptation)} \cite{chen2018femosaa}: The approach responds to workload changes by triggering a new search from scratch with randomly initialized configurations.
	
	\item \textbf{\texttt{Seed-EA} (Dynamic Adaptation)} \cite{kinneer2021information}: By using an evolutionary algorithm, the approach seeds all the configurations preserved from the most recent past workload for planning under the new workload.
	
	\item \textbf{\texttt{D-SOGA} (Mixed Adaptation):} As a single-objective variant of D-NSGA-II \cite{deb2007dynamic}, this approach retains 80\% randomly chosen configurations from the most recent past workload with 20\% new randomly initialized configurations to preserve diversity when the workload change.
	
	\item \textbf{\texttt{LiDOS} (Dynamic Adaptation)} \cite{chen2022lifelong}: The approach transforms single-objective problems into multi-objective ones via an auxiliary objective, thereby leveraging non-dominance relations to retain local optimal configurations under the most recent past workload to be seeded for the new workload.
\end{itemize}

\subsection{Component and Parameter Settings}
\label{subsection:component_and_parameter_settings}
For a fair comparison across all approaches, the parameters of all stochastic search algorithms in the planning are standardized, where binary tournament is employed for mating selection, together with the boundary mutation and single-point crossover. The mutation and crossover rates are set at 0.1 and 0.9, respectively, with a population size of 20, which is widely used in prior works~\cite{chen2018femosaa,chen2021mmo}. For \approach, we set its only parameter $\alpha=0.3$, unless otherwise stated, as this tends to be the generally best setting (see Section \ref{subsection:sensitivity_to_alpha}). 

\revision{In this study, we use Cyber-Twin to mimic the behaviors of the managed systems, which aims to expedite configuration evaluation in planning by using less time/resources without interfering with the managed system. There are different ways to create such a Cyber-Twin~\cite{eramo2021conceptualizing}, e.g., (1) building a data driven surrogate model; (2) using existing benchmarks; or (3) creating a low-cost simulator/replica. Here, we chose existing benchmarks from \cite{muhlbauer2023analyzing} as the Cyber-Twin for all the compared adaptation approaches in experiments, which is straightforward and easy to implement, providing reliable performance data for accurate configuration measurements \cite{chen2022lifelong,DBLP:journals/tse/Nair0MSA20}.} Particularly, the budget constraint is 80 measurements (i.e., $R_t=80$), which is sufficient for the approaches to converge while allowing them to timely self-adapt the system. The achieved performance at the end of planning is recorded. For the purpose of a controlled experiment, the workload would change shortly after the best configuration from the previous planning has been used to adapt the system.


In the search/optimization procedure of the planning under a particular workload, for all approaches, we do not explore duplicate configurations, i.e., only the newly evaluated/measured configurations from the Cyber-Twin would consume the budget. When the planning finds invalid configurations, we give a purposely worsened performance to those configurations, hence they would naturally be ruled out during the search/optimization process.



To obtain a statistically sound comparison, all experiments in this study are run 100 times independently. In particular, \revision{to test the self-adaptation of the systems, each of the runs follows a randomly shuffled order of the workloads, allowing us to alleviate the bias introduced by a specific occurrence sequence of the workloads.} The performance results under each workload are used but it might appear at a different position across the repeated runs.

\subsection{Statistical Validation}
We employed different statistical validation for examining the 100 runs of results.

\subsubsection{Pairwise Comparisons} For this, we use the following:

\begin{itemize}
	\item \textbf{Non-parametric test:} We use Wilcoxon rank-sum test---a common test widely used in SBSE for its strong statistical power on pairwise comparisons \cite{arcuri2011practical}. The significance level over 100 runs is set at 0.05 and $p < 0.05$ indicates significant performance differences in the comparison.
	
	\item \textbf{Effect size:} In addition to statistical significance, we use $\hat{A}_{12}$ to measure the effect size~\cite{vargha2000critique}.  Particularly, $\hat{A}_{12} \leq 0.44$ or $\hat{A}_{12} \geq 0.56$ suggests a non-trivial effect.
 
\end{itemize}

Thus, we say a comparison is statistically significant only if it has $\hat{A}_{12} \geq 0.56$  (or $\hat{A}_{12} \leq 0.44$) and $p < 0.05$.

\begin{table*}[t!]
	\setlength{\tabcolsep}{0.8mm}
	\caption{The Mean and Standard deviation (Std) of performance objectives between \approach~and other state-of-the-art approaches for all cases over 100 runs. For each case, \setlength{\fboxsep}{1.5pt}\colorbox{teal!30}{green cells} mean \approach~has the best mean performance; or \setlength{\fboxsep}{1.5pt}\colorbox{red!20}{red cells} otherwise. The one(s) with the best rank ($r$) from the Scott-Knott test is highlighted in bold.}
 \label{sec:rq1}
	\begin{adjustbox}{width=\textwidth,center}
		\resizebox{\textwidth}{!}{ 
	\begin{tabular}{>{\centering\arraybackslash}p{1.35cm}|p{1.35cm}|p{0.3cm}p{2.0cm}|p{0.3cm}p{1.95cm}|p{0.3cm}p{1.95cm}|p{0.3cm}p{1.95cm}|p{0.3cm}p{1.95cm}|p{0.3cm}p{1.95cm}|p{0.3cm}p{1.95cm}|p{0.3cm}p{2.8cm}|p{0.3cm}p{1.95cm}}
		\hline	
 &
   &
  \multicolumn{2}{c|}{\textbf{\textsc{lrzip}}} &
  \multicolumn{2}{c|}{\textbf{\textsc{xz}}} &
  \multicolumn{2}{c|}{\textbf{\textsc{z3}}} &
  \multicolumn{2}{c|}{\textbf{\textsc{dconvert}}} &
  \multicolumn{2}{c|}{\textbf{\textsc{batlik}}} &
  \multicolumn{2}{c|}{\textbf{\textsc{kanzi}}} &
  \multicolumn{2}{c|}{\textbf{\textsc{x264}}} &
  \multicolumn{2}{c|}{\textbf{\textsc{h2}}} &
  \multicolumn{2}{c}{\textbf{\textsc{jump3r}}} \\ \cline{3-20} 
\multirow{-2}{*}{\textbf{Workload}} &
  \multirow{-2}{*}{\textbf{Approach}} &
  \textit{\textbf{\bm{$r$}}} &
  \textbf{Mean (Std)} &
  \textit{\textbf{\bm{$r$}}} &
  \textbf{Mean (Std)} &
  \textit{\textbf{\bm{$r$}}} &
  \textbf{Mean (Std)} &
  \textit{\textbf{\bm{$r$}}} &
  \textbf{Mean (Std)} &
  \textit{\textbf{\bm{$r$}}} &
  \textbf{Mean (Std)} &
  \textit{\textbf{\bm{$r$}}} &
  \textbf{Mean (Std)} &
  \textit{\textbf{\bm{$r$}}} &
  \textbf{Mean (Std)} &
  \textit{\textbf{\bm{$r$}}} &
  \textbf{Mean (Std)} &
  \textit{\textbf{\bm{$r$}}} &
  \textbf{Mean (Std)} \\ \hline
 &
  FEMOSAA &
  2 &
  3.214 (0.127) &
  3 &
  4.931 (1.445) &
  \textbf{1} &
  \textbf{5.881 (0.131)} &
  4 &
  1.952 (0.147) &
  3 &
  0.953 (0.043) &
  3 &
  1.830 (1.330) &
  3 &
  1.038 (0.299) &
  3 &
  25641.195 (1612.807) &
  2 &
  2.979 (1.030) \\
 &
  Seed-EA &
  2 &
  3.134 (0.034) &
  2 &
  4.636 (1.427) &
  2 &
  5.898 (0.317) &
  3 &
  1.867 (0.124) &
  2 &
  0.912 (0.031) &
  2 &
  1.346 (1.286) &
  2 &
  0.937 (0.190) &
  2 &
  26395.457 (1332.090) &
  2 &
  2.636 (0.836) \\
 &
  D-SOGA &
  \cellcolor{red!20}\textbf{1} &
  \cellcolor{red!20}\textbf{3.132 (0.015)} &
  \textbf{1} &
  \textbf{4.486 (1.251)} &
  2 &
  5.951 (0.953) &
  \cellcolor{red!20}\textbf{1} &
  \cellcolor{red!20}\textbf{1.844 (0.093)} &
  2 &
  0.912 (0.024) &
  2 &
  1.252 (1.078) &
  2 &
  0.933 (0.189) &
  2 &
  26485.351 (1090.741) &
  2 &
  2.666 (0.848) \\
 &
  LiDOS &
  2 &
  3.136 (0.032) &
  2 &
  4.924 (1.986) &
  3 &
  5.993 (0.628) &
  4 &
  1.892 (0.135) &
  2 &
  0.916 (0.025) &
  2 &
  1.386 (1.252) &
  2 &
  0.965 (0.195) &
  2 &
  26287.284 (1466.595) &
  2 &
  2.665 (0.789) \\
\multirow{-5}{*}{\textbf{W1}} &
  DLiSA &
  2 &
  3.135 (0.035) &
  \cellcolor{teal!30}\textbf{1} &
  \cellcolor{teal!30}\textbf{3.813 (0.849)} &
  \cellcolor{teal!30}\textbf{1} &
  \cellcolor{teal!30}\textbf{5.856 (0.011)} &
  2 &
  1.849 (0.105) &
  \cellcolor{teal!30}\textbf{1} &
  \cellcolor{teal!30}\textbf{0.907 (0.014)} &
  \cellcolor{teal!30}\textbf{1} &
  \cellcolor{teal!30}\textbf{0.986 (0.866)} &
  \cellcolor{teal!30}\textbf{1} &
  \cellcolor{teal!30}\textbf{0.890 (0.140)} &
  \cellcolor{teal!30}\textbf{1} &
  \cellcolor{teal!30}\textbf{26721.450 (705.601)} &
  \cellcolor{teal!30}\textbf{1} &
  \cellcolor{teal!30}\textbf{2.573 (0.828)} \\ \hline
 &
  FEMOSAA &
  2 &
  0.031 (0.002) &
  2 &
  0.014 (0.005) &
  \cellcolor{red!20}\textbf{1} &
  \cellcolor{red!20}\textbf{1.77 (0.089)} &
  2 &
  1.186 (0.071) &
  3 &
  1.38 (0.037) &
  3 &
  0.166 (0.048) &
  3 &
  3.954 (0.842) &
  3 &
  18273.524 (1026.898) &
  3 &
  1.093 (0.314) \\
 &
  Seed-EA &
  2 &
  0.030 (0.000) &
  2 &
  0.013 (0.005) &
  4 &
  2.463 (0.668) &
  2 &
  1.12 (0.061) &
  2 &
  1.340 (0.020) &
  2 &
  0.143 (0.041) &
  2 &
  3.775 (0.806) &
  2 &
  18573.352 (1625.003) &
  2 &
  0.915 (0.279) \\
 &
  D-SOGA &
  2 &
  0.030 (0.000) &
  2 &
  0.013 (0.005) &
  \textbf{1} &
  \textbf{2.051 (0.491)} &
  2 &
  1.118 (0.056) &
  3 &
  1.341 (0.019) &
  2 &
  0.141 (0.033) &
  \textbf{1} &
  \textbf{3.751 (0.819)} &
  \textbf{1} &
  \textbf{18847.953 (865.318)} &
  2 &
  0.944 (0.264) \\
 &
  LiDOS &
  2 &
  0.030 (0.000) &
  3 &
  0.015 (0.006) &
  3 &
  2.375 (0.616) &
  2 &
  1.126 (0.062) &
  3 &
  1.345 (0.028) &
  2 &
  0.147 (0.045) &
  2 &
  3.927 (0.845) &
  2 &
  18488.896 (1651.846) &
  2 &
  0.915 (0.232) \\
\multirow{-5}{*}{\textbf{W2}} &
  DLiSA &
  \cellcolor{teal!30}\textbf{1} &
  \cellcolor{teal!30}\textbf{0.030 (0.000)} &
  \cellcolor{teal!30}\textbf{1} &
  \cellcolor{teal!30}\textbf{0.011 (0.003)} &
  2 &
  2.254 (0.608) &
  \cellcolor{teal!30}\textbf{1} &
  \cellcolor{teal!30}\textbf{1.115 (0.049)} &
  \cellcolor{teal!30}\textbf{1} &
  \cellcolor{teal!30}\textbf{1.338 (0.019)} &
  \cellcolor{teal!30}\textbf{1} &
  \cellcolor{teal!30}\textbf{0.131 (0.032)} &
  \cellcolor{teal!30}\textbf{1} &
  \cellcolor{teal!30}\textbf{3.590 (0.567)} &
  \cellcolor{teal!30}\textbf{1} &
  \cellcolor{teal!30}\textbf{18972.982 (758.262)} &
  \cellcolor{teal!30}\textbf{1} &
  \cellcolor{teal!30}\textbf{0.846 (0.197)} \\ \hline
 &
  FEMOSAA &
  3 &
  3.335 (0.037) &
  2 &
  4.804 (1.461) &
  2 &
  0.629 (0.999) &
  3 &
  0.384 (0.007) &
  2 &
  4.326 (0.153) &
  \cellcolor{red!20}\textbf{1} &
  \cellcolor{red!20}\textbf{0.293 (0.12)} &
  3 &
  1.534 (0.421) &
  3 &
  908.110 (46.396) &
  3 &
  1.629 (0.514) \\
 &
  Seed-EA &
  \cellcolor{red!20}\textbf{1} &
  \cellcolor{red!20}\textbf{3.304 (0.013)} &
  \textbf{1} &
  \textbf{4.558 (1.412)} &
  2 &
  0.417 (0.802) &
  2 &
  0.375 (0.007) &
  2 &
  4.197 (0.040) &
  3 &
  0.458 (0.821) &
  2 &
  1.374 (0.327) &
  \textbf{1} &
  \textbf{943.275 (53.605)} &
  2 &
  1.375 (0.421) \\
 &
  D-SOGA &
  3 &
  3.311 (0.023) &
  \textbf{1} &
  \textbf{4.543 (1.394)} &
  \cellcolor{red!20}\textbf{1} &
  \cellcolor{red!20}\textbf{0.352 (0.611)} &
  3 &
  0.376 (0.007) &
  2 &
  4.199 (0.042) &
  3 &
  0.322 (0.168) &
  2 &
  1.400 (0.322) &
  \textbf{1} &
  \textbf{946.655 (43.526)} &
  3 &
  1.441 (0.414) \\
 &
  LiDOS &
  2 &
  3.309 (0.019) &
  \textbf{1} &
  \textbf{4.641 (1.438)} &
  2 &
  0.429 (0.875) &
  3 &
  0.377 (0.008) &
  2 &
  4.206 (0.056) &
  3 &
  0.659 (0.706) &
  2 &
  1.422 (0.348) &
  2 &
  938.221 (54.261) &
  2 &
  1.431 (0.372) \\
\multirow{-5}{*}{\textbf{W3}} &
  DLiSA &
  2 &
  3.305 (0.014) &
  \cellcolor{teal!30}\textbf{1} &
  \cellcolor{teal!30}\textbf{3.835 (0.966)} &
  2 &
  0.364 (0.660) &
  \cellcolor{teal!30}\textbf{1} &
  \cellcolor{teal!30}\textbf{0.375 (0.008)} &
  \cellcolor{teal!30}\textbf{1} &
  \cellcolor{teal!30}\textbf{4.196 (0.056)} &
  2 &
  0.308 (0.129) &
  \cellcolor{teal!30}\textbf{1} &
  \cellcolor{teal!30}\textbf{1.286 (0.248)} &
  \cellcolor{teal!30}\textbf{1} &
  \cellcolor{teal!30}\textbf{948.344 (38.602)} &
  \cellcolor{teal!30}\textbf{1} &
  \cellcolor{teal!30}\textbf{1.309 (0.368)} \\ \hline
 &
  FEMOSAA &
  3 &
  7.268 (0.227) &
  \textbf{1} &
  \textbf{13.679 (3.906)} &
  \textbf{1} &
  \textbf{2.375 (0.245)} &
  3 &
  1.657 (0.081) &
  3 &
  1.233 (0.030) &
  4 &
  2.625 (1.845) &
  3 &
  1.770 (0.438) &
  3 &
  963.053 (92.521) &
  3 &
  0.741 (0.157) \\
 &
  Seed-EA &
  2 &
  7.165 (0.088) &
  \textbf{1} &
  \textbf{13.133 (3.830)} &
  2 &
  2.422 (0.274) &
  3 &
  1.608 (0.071) &
  \cellcolor{red!20}\textbf{1} &
  \cellcolor{red!20}\textbf{1.191 (0.022)} &
  2 &
  1.851 (1.789) &
  \textbf{1} &
  \textbf{1.702 (0.430)} &
  2 &
  1012.475 (108.587) &
  2 &
  0.685 (0.145) \\
 &
  D-SOGA &
  2 &
  7.170 (0.078) &
  \textbf{1} &
  \textbf{13.132 (4.012)} &
  3 &
  2.468 (0.428) &
  \cellcolor{red!20}\textbf{1} &
  \cellcolor{red!20}\textbf{1.603 (0.063)} &
  3 &
  1.195 (0.023) &
  2 &
  1.598 (1.403) &
  \textbf{1} &
  \textbf{1.652 (0.347)} &
  2 &
  1015.531 (86.929) &
  2 &
  0.697 (0.136) \\
 &
  LiDOS &
  3 &
  7.171 (0.039) &
  2 &
  14.171 (8.349) &
  2 &
  2.384 (0.254) &
  3 &
  1.622 (0.08) &
  3 &
  1.200 (0.021) &
  3 &
  1.963 (1.883) &
  2 &
  1.757 (0.413) &
  2 &
  1006.740 (104.145) &
  2 &
  0.679 (0.105) \\
\multirow{-5}{*}{\textbf{W4}} &
  DLiSA &
  \cellcolor{teal!30}\textbf{1} &
  \cellcolor{teal!30}\textbf{7.159 (0.032)} &
  \cellcolor{teal!30}\textbf{1} &
  \cellcolor{teal!30}\textbf{11.102 (2.730)} &
  \cellcolor{teal!30}\textbf{1} &
  \cellcolor{teal!30}\textbf{2.324 (0.150)} &
  2 &
  1.605 (0.067) &
  2 &
  1.193 (0.026) &
  \cellcolor{teal!30}\textbf{1} &
  \cellcolor{teal!30}\textbf{1.173 (0.697)} &
  \cellcolor{teal!30}\textbf{1} &
  \cellcolor{teal!30}\textbf{1.586 (0.236)} &
  \cellcolor{teal!30}\textbf{1} &
  \cellcolor{teal!30}\textbf{1032.006 (45.261)} &
  \cellcolor{teal!30}\textbf{1} &
  \cellcolor{teal!30}\textbf{0.642 (0.076)} \\ \hline
 &
  FEMOSAA &
  3 &
  33.581 (0.386) &
  2 &
  14.266 (3.910) &
  2 &
  3.339 (0.655) &
  2 &
  0.522 (0.021) &
  3 &
  2.492 (0.066) &
  3 &
  1.884 (1.057) &
  3 &
  4.073 (2.727) &
  2 &
  46866.246 (3439.525) &
  3 &
  1.446 (0.699) \\
 &
  Seed-EA &
  \cellcolor{red!20}\textbf{1} &
  \cellcolor{red!20}\textbf{33.395 (0.016)} &
  \textbf{1} &
  \textbf{13.657 (4.411)} &
  2 &
  3.180 (0.330) &
  \cellcolor{red!20}\textbf{1} &
  \cellcolor{red!20}\textbf{0.502 (0.015)} &
  2 &
  2.409 (0.040) &
  2 &
  1.281 (1.076) &
  2 &
  3.414 (0.729) &
  \textbf{1} &
  \textbf{47332.765 (3793.023)} &
  2 &
  1.136 (0.417) \\
 &
  D-SOGA &
  2 &
  33.397 (0.017) &
  \textbf{1} &
  \textbf{13.815 (3.875)} &
  2 &
  3.172 (0.223) &
  2 &
  0.503 (0.015) &
  2 &
  2.413 (0.042) &
  2 &
  1.182 (0.994) &
  2 &
  3.438 (0.795) &
  \textbf{1} &
  \textbf{47491.315 (3459.523)} &
  2 &
  1.210 (0.553) \\
 &
  LiDOS &
  3 &
  33.424 (0.148) &
  \textbf{1} &
  \textbf{14.018 (4.555)} &
  2 &
  3.195 (0.342) &
  2 &
  0.505 (0.016) &
  3 &
  2.420 (0.041) &
  2 &
  1.362 (1.152) &
  2 &
  3.530 (0.808) &
  2 &
  47021.762 (4273.182) &
  2 &
  1.179 (0.44) \\
\multirow{-5}{*}{\textbf{W5}} &
  DLiSA &
  2 &
  33.421 (0.150) &
  \cellcolor{teal!30}\textbf{1} &
  \cellcolor{teal!30}\textbf{11.702 (3.297)} &
  \cellcolor{teal!30}\textbf{1} &
  \cellcolor{teal!30}\textbf{3.150 (0.111)} &
  2 &
  0.503 (0.019) &
  \cellcolor{teal!30}\textbf{1} &
  \cellcolor{teal!30}\textbf{2.404 (0.036)} &
  \cellcolor{teal!30}\textbf{1} &
  \cellcolor{teal!30}\textbf{0.938 (0.604)} &
  \cellcolor{teal!30}\textbf{1} &
  \cellcolor{teal!30}\textbf{3.222 (0.514)} &
  \cellcolor{teal!30}\textbf{1} &
  \cellcolor{teal!30}\textbf{47835.194 (2491.758)} &
  \cellcolor{teal!30}\textbf{1} &
  \cellcolor{teal!30}\textbf{1.045 (0.246)} \\ \hline
 &
  FEMOSAA &
  4 &
  0.978 (0.012) &
  3 &
  2.172 (0.634) &
  2 &
  1.406 (0.228) &
  2 &
  0.391 (0.013) &
  3 &
  3.323 (0.157) &
  4 &
  0.687 (0.448) &
  3 &
  0.111 (0.018) &
  3 &
  47199.317 (2383.084) &
  4 &
  0.319 (0.045) \\
 &
  Seed-EA &
  2 &
  0.971 (0.003) &
  \textbf{1} &
  \textbf{1.926 (0.513)} &
  2 &
  1.330 (0.135) &
  \cellcolor{red!20}\textbf{1} &
  \cellcolor{red!20}\textbf{0.375 (0.011)} &
  2 &
  3.158 (0.059) &
  3 &
  0.528 (0.406) &
  2 &
  0.104 (0.014) &
  3 &
  47446.104 (3798.389) &
  2 &
  0.304 (0.028) \\
 &
  D-SOGA &
  \cellcolor{red!20}\textbf{1} &
  \cellcolor{red!20}\textbf{0.97 (0.002)} &
  2 &
  2.077 (0.696) &
  2 &
  1.327 (0.137) &
  2 &
  0.376 (0.012) &
  3 &
  3.158 (0.049) &
  2 &
  0.519 (0.363) &
  2 &
  0.103 (0.012) &
  2 &
  47844.701 (2876.854) &
  2 &
  0.309 (0.033) \\
 &
  LiDOS &
  3 &
  0.972 (0.006) &
  \textbf{1} &
  \textbf{2.053 (0.724)} &
  2 &
  1.337 (0.158) &
  2 &
  0.376 (0.01) &
  3 &
  3.170 (0.049) &
  3 &
  0.581 (0.440) &
  2 &
  0.105 (0.015) &
  3 &
  47119.407 (4400.199) &
  3 &
  0.310 (0.030) \\
\multirow{-5}{*}{\textbf{W6}} &
  DLiSA &
  3 &
  0.971 (0.003) &
  \cellcolor{teal!30}\textbf{1} &
  \cellcolor{teal!30}\textbf{1.638 (0.375)} &
  \cellcolor{teal!30}\textbf{1} &
  \cellcolor{teal!30}\textbf{1.322 (0.130)} &
  2 &
  0.376 (0.011) &
  \cellcolor{teal!30}\textbf{1} &
  \cellcolor{teal!30}\textbf{3.152 (0.042)} &
  \cellcolor{teal!30}\textbf{1} &
  \cellcolor{teal!30}\textbf{0.433 (0.263)} &
  \cellcolor{teal!30}\textbf{1} &
  \cellcolor{teal!30}\textbf{0.100 (0.013)} &
  \cellcolor{teal!30}\textbf{1} &
  \cellcolor{teal!30}\textbf{48335.083 (488.968)} &
  \cellcolor{teal!30}\textbf{1} &
  \cellcolor{teal!30}\textbf{0.298 (0.018)} \\ \hline
 &
  FEMOSAA &
  3 &
  0.198 (0.005) &
  3 &
  0.215 (0.023) &
  \textbf{1} &
  \textbf{0.278 (0.212)} &
  2 &
  19.015 (3.692) &
  2 &
  1.170 (0.036) &
  3 &
  0.243 (0.106) &
  3 &
  0.679 (0.206) &
  2 &
  18499.348 (1805.899) &
  \multicolumn{2}{c}{} \\
 &
  Seed-EA &
  \cellcolor{red!20}\textbf{1} &
  \cellcolor{red!20}\textbf{0.192 (0.004)} &
  \textbf{1} &
  \textbf{0.205 (0.023)} &
  \textbf{1} &
  \textbf{0.320 (0.504)} &
  2 &
  17.485 (2.864) &
  \cellcolor{red!20}\textbf{1} &
  \cellcolor{red!20}\textbf{1.136 (0.016)} &
  2 &
  0.210 (0.108) &
  2 &
  0.606 (0.154) &
  \textbf{1} &
  \textbf{19979.516 (1662.321)} &
  \multicolumn{2}{c}{} \\
 &
  D-SOGA &
  2 &
  0.192 (0.004) &
  2 &
  0.206 (0.019) &
  \cellcolor{red!20}\textbf{1} &
  \cellcolor{red!20}\textbf{0.265 (0.194)} &
  \cellcolor{red!20}\textbf{1} &
  \cellcolor{red!20}\textbf{17.269 (2.668)} &
  2 &
  1.136 (0.013) &
  2 &
  0.205 (0.104) &
  2 &
  0.606 (0.161) &
  \textbf{1} &
  \textbf{19952.252 (1608.535)} &
  \multicolumn{2}{c}{} \\
 &
  LiDOS &
  3 &
  0.193 (0.005) &
  2 &
  0.214 (0.031) &
  \textbf{1} &
  \textbf{0.269 (0.253)} &
  2 &
  18.333 (3.433) &
  2 &
  1.139 (0.022) &
  2 &
  0.212 (0.116) &
  2 &
  0.632 (0.173) &
  \textbf{1} &
  \textbf{19878.876 (1701.235)} &
  \multicolumn{2}{c}{} \\
\multirow{-5}{*}{\textbf{W7}} &
  DLiSA &
  3 &
  0.192 (0.004) &
  \cellcolor{teal!30}\textbf{1} &
  \cellcolor{teal!30}\textbf{0.196 (0.015)} &
  \textbf{1} &
  \textbf{0.292 (0.458)} &
  \textbf{1} &
  \textbf{17.366 (2.734)} &
  2 &
  1.137 (0.016) &
  \cellcolor{teal!30}\textbf{1} &
  \cellcolor{teal!30}\textbf{0.177 (0.078)} &
  \cellcolor{teal!30}\textbf{1} &
  \cellcolor{teal!30}\textbf{0.572 (0.110)} &
  \cellcolor{teal!30}\textbf{1} &
  \cellcolor{teal!30}\textbf{20037.040 (1584.735)} &
  \multicolumn{2}{c}{\multirow{-5}{*}{N/A}} \\ \hline
 &
  FEMOSAA &
  3 &
  10.966 (0.092) &
  \textbf{1} &
  \textbf{29.552 (8.366)} &
  3 &
  8.751 (0.016) &
  3 &
  1.057 (0.033) &
  3 &
  7.249 (0.251) &
  4 &
  5.386 (4.813) &
  3 &
  0.162 (0.086) &
  2 &
  26235.578 (2089.717) &
  \multicolumn{2}{c}{} \\
 &
  Seed-EA &
  3 &
  10.910 (0.043) &
  \textbf{1} &
  \textbf{28.616 (8.807)} &
  3 &
  8.747 (0.015) &
  \cellcolor{red!20}\textbf{1} &
  \cellcolor{red!20}\textbf{1.030 (0.026)} &
  2 &
  7.079 (0.112) &
  2 &
  3.869 (4.276) &
  2 &
  0.142 (0.029) &
  \textbf{1} &
  \textbf{27973.092 (2149.536)} &
  \multicolumn{2}{c}{} \\
 &
  D-SOGA &
  2 &
  10.909 (0.039) &
  \textbf{1} &
  \textbf{28.969 (8.508)} &
  3 &
  8.806 (0.590) &
  2 &
  1.032 (0.026) &
  3 &
  7.084 (0.105) &
  2 &
  3.227 (2.843) &
  2 &
  0.143 (0.025) &
  \cellcolor{red!20}\textbf{1} &
  \cellcolor{red!20}\textbf{28147.533 (1915.831)} &
  \multicolumn{2}{c}{} \\
 &
  LiDOS &
  3 &
  10.919 (0.048) &
  2 &
  29.977 (9.191) &
  \cellcolor{red!20}\textbf{1} &
  \cellcolor{red!20}\textbf{8.746 (0.005)} &
  3 &
  1.039 (0.029) &
  3 &
  7.095 (0.093) &
  3 &
  4.002 (4.353) &
  2 &
  0.146 (0.032) &
  \textbf{1} &
  \textbf{27926.014 (2064.011)} &
  \multicolumn{2}{c}{} \\
\multirow{-5}{*}{\textbf{W8}} &
  DLiSA &
  \cellcolor{teal!30}\textbf{1} &
  \cellcolor{teal!30}\textbf{10.907 (0.020)} &
  \cellcolor{teal!30}\textbf{1} &
  \cellcolor{teal!30}\textbf{23.789 (5.998)} &
  2 &
  8.746 (0.005) &
  3 &
  1.032 (0.027) &
  \cellcolor{teal!30}\textbf{1} &
  \cellcolor{teal!30}\textbf{7.076 (0.077)} &
  \cellcolor{teal!30}\textbf{1} &
  \cellcolor{teal!30}\textbf{2.347 (2.228)} &
  \cellcolor{teal!30}\textbf{1} &
  \cellcolor{teal!30}\textbf{0.133 (0.019)} &
  \textbf{1} &
  \textbf{28129.890 (1669.565)} &
  \multicolumn{2}{c}{\multirow{-5}{*}{N/A}} \\ \hline
 &
  FEMOSAA &
  4 &
  9.492 (0.478) &
  \textbf{1} &
  \textbf{27.128 (7.707)} &
  3 &
  3.184 (0.005) &
  3 &
  0.491 (0.015) &
  3 &
  1.068 (0.018) &
  3 &
  1.092 (0.695) &
  3 &
  0.261 (0.050) &
  \multicolumn{2}{c|}{} &
  \multicolumn{2}{c}{} \\
 &
  Seed-EA &
  \cellcolor{red!20}\textbf{1} &
  \cellcolor{red!20}\textbf{9.180 (0.282)} &
  \textbf{1} &
  \textbf{25.640 (7.097)} &
  2 &
  3.181 (0.003) &
  2 &
  0.474 (0.015) &
  \cellcolor{red!20}\textbf{1} &
  \cellcolor{red!20}\textbf{1.047 (0.01)} &
  2 &
  0.852 (0.615) &
  \textbf{1} &
  \textbf{0.249 (0.038)} &
  \multicolumn{2}{c|}{} &
  \multicolumn{2}{c}{} \\
 &
  D-SOGA &
  3 &
  9.279 (0.321) &
  \textbf{1} &
  \textbf{26.330 (7.248)} &
  \cellcolor{red!20}\textbf{1} &
  \cellcolor{red!20}\textbf{3.181 (0.003)} &
  3 &
  0.474 (0.014) &
  2 &
  1.049 (0.012) &
  2 &
  0.813 (0.583) &
  2 &
  0.249 (0.039) &
  \multicolumn{2}{c|}{} &
  \multicolumn{2}{c}{} \\
 &
  LiDOS &
  4 &
  9.358 (0.296) &
  2 &
  27.838 (16.49) &
  3 &
  3.182 (0.004) &
  3 &
  0.477 (0.016) &
  2 &
  1.049 (0.012) &
  2 &
  0.912 (0.745) &
  2 &
  0.253 (0.040) &
  \multicolumn{2}{c|}{} &
  \multicolumn{2}{c}{} \\
\multirow{-5}{*}{\textbf{W9}} &
  DLiSA &
  2 &
  9.197 (0.314) &
  \cellcolor{teal!30}\textbf{1} &
  \cellcolor{teal!30}\textbf{21.324 (5.188)} &
  3 &
  3.181 (0.003) &
  \cellcolor{teal!30}\textbf{1} &
  \cellcolor{teal!30}\textbf{0.473 (0.014)} &
  2 &
  1.051 (0.014) &
  \cellcolor{teal!30}\textbf{1} &
  \cellcolor{teal!30}\textbf{0.709 (0.585)} &
  \cellcolor{teal!30}\textbf{1} &
  \cellcolor{teal!30}\textbf{0.240 (0.031)} &
  \multicolumn{2}{c|}{\multirow{-5}{*}{N/A}} &
  \multicolumn{2}{c}{\multirow{-5}{*}{N/A}} \\ \hline
 &
  FEMOSAA &
  3 &
  5.595 (0.381) &
  \textbf{1} &
  \textbf{13.081 (3.964)} &
  \cellcolor{red!20}\textbf{1} &
  \cellcolor{red!20}\textbf{6.795 (0.232)} &
  2 &
  1.441 (0.011) &
  3 &
  1.145 (0.037) &
  \multicolumn{2}{c|}{} &
  \multicolumn{2}{c|}{} &
  \multicolumn{2}{c|}{} &
  \multicolumn{2}{c}{} \\
 &
  Seed-EA &
  2 &
  5.358 (0.233) &
  \textbf{1} &
  \textbf{12.952 (4.033)} &
  \textbf{1} &
  \textbf{6.822 (0.233)} &
  \textbf{1} &
  \textbf{1.439 (0.009)} &
  \cellcolor{red!20}\textbf{1} &
  \cellcolor{red!20}\textbf{1.116 (0.015)} &
  \multicolumn{2}{c|}{} &
  \multicolumn{2}{c|}{} &
  \multicolumn{2}{c|}{} &
  \multicolumn{2}{c}{} \\
 &
  D-SOGA &
  3 &
  5.429 (0.299) &
  \textbf{1} &
  \textbf{12.878 (3.446)} &
  \textbf{1} &
  \textbf{6.840 (0.241)} &
  \textbf{1} &
  \textbf{1.44 (0.008)} &
  3 &
  1.117 (0.017) &
  \multicolumn{2}{c|}{} &
  \multicolumn{2}{c|}{} &
  \multicolumn{2}{c|}{} &
  \multicolumn{2}{c}{} \\
 &
  LiDOS &
  2 &
  5.398 (0.276) &
  2 &
  13.191 (4.421) &
  \textbf{1} &
  \textbf{6.845 (0.265)} &
  2 &
  1.441 (0.009) &
  3 &
  1.120 (0.018) &
  \multicolumn{2}{c|}{} &
  \multicolumn{2}{c|}{} &
  \multicolumn{2}{c|}{} &
  \multicolumn{2}{c}{} \\
\multirow{-5}{*}{\textbf{W10}} &
  DLiSA &
  \cellcolor{teal!30}\textbf{1} &
  \cellcolor{teal!30}\textbf{5.358 (0.228)} &
  \cellcolor{teal!30}\textbf{1} &
  \cellcolor{teal!30}\textbf{10.605 (2.606)} &
  \textbf{1} &
  \textbf{6.816 (0.236)} &
  \cellcolor{teal!30}\textbf{1} &
  \cellcolor{teal!30}\textbf{1.438 (0.009)} &
  2 &
  1.117 (0.017) &
  \multicolumn{2}{c|}{\multirow{-5}{*}{N/A}} &
  \multicolumn{2}{c|}{\multirow{-5}{*}{N/A}} &
  \multicolumn{2}{c|}{\multirow{-5}{*}{N/A}} &
  \multicolumn{2}{c}{\multirow{-5}{*}{N/A}} \\ \hline
 &
  FEMOSAA &
  2 &
  2.126 (0.047) &
  3 &
  3.388 (0.990) &
  \textbf{1} &
  \textbf{8.248 (1.047)} &
  3 &
  1.464 (0.02) &
  3 &
  1.693 (0.053) &
  \multicolumn{2}{c|}{} &
  \multicolumn{2}{c|}{} &
  \multicolumn{2}{c|}{} &
  \multicolumn{2}{c}{} \\
 &
  Seed-EA &
  2 &
  2.092 (0.035) &
  2 &
  3.254 (0.913) &
  2 &
  8.410 (1.719) &
  \cellcolor{red!20}\textbf{1} &
  \cellcolor{red!20}\textbf{1.442 (0.018)} &
  \cellcolor{red!20}\textbf{1} &
  \cellcolor{red!20}\textbf{1.627 (0.035)} &
  \multicolumn{2}{c|}{} &
  \multicolumn{2}{c|}{} &
  \multicolumn{2}{c|}{} &
  \multicolumn{2}{c}{} \\
 &
  D-SOGA &
  2 &
  2.094 (0.029) &
  \textbf{1} &
  \textbf{3.190 (0.943)} &
  \textbf{1} &
  \textbf{8.232 (1.282)} &
  2 &
  1.446 (0.019) &
  3 &
  1.630 (0.032) &
  \multicolumn{2}{c|}{} &
  \multicolumn{2}{c|}{} &
  \multicolumn{2}{c|}{} &
  \multicolumn{2}{c}{} \\
 &
  LiDOS &
  2 &
  2.093 (0.029) &
  2 &
  3.311 (0.962) &
  \textbf{1} &
  \textbf{8.327 (1.378)} &
  3 &
  1.447 (0.021) &
  3 &
  1.636 (0.041) &
  \multicolumn{2}{c|}{} &
  \multicolumn{2}{c|}{} &
  \multicolumn{2}{c|}{} &
  \multicolumn{2}{c}{} \\
\multirow{-5}{*}{\textbf{W11}} &
  DLiSA &
  \cellcolor{teal!30}\textbf{1} &
  \cellcolor{teal!30}\textbf{2.089 (0.022)} &
  \cellcolor{teal!30}\textbf{1} &
  \cellcolor{teal!30}\textbf{2.804 (0.775)} &
  \cellcolor{teal!30}\textbf{1} &
  \cellcolor{teal!30}\textbf{7.948 (0.654)} &
  2 &
  1.444 (0.019) &
  2 &
  1.628 (0.038) &
  \multicolumn{2}{c|}{\multirow{-5}{*}{N/A}} &
  \multicolumn{2}{c|}{\multirow{-5}{*}{N/A}} &
  \multicolumn{2}{c|}{\multirow{-5}{*}{N/A}} &
  \multicolumn{2}{c}{\multirow{-5}{*}{N/A}} \\ \hline
 &
  FEMOSAA &
  3 &
  3.547 (0.121) &
  3 &
  7.159 (2.183) &
  3 &
  3.989 (0.344) &
  2 &
  0.49 (0.009) &
  \multicolumn{2}{c|}{} &
  \multicolumn{2}{c|}{} &
  \multicolumn{2}{c|}{} &
  \multicolumn{2}{c|}{} &
  \multicolumn{2}{c}{} \\
 &
  Seed-EA &
  \cellcolor{red!20}\textbf{1} &
  \cellcolor{red!20}\textbf{3.474 (0.059)} &
  2 &
  6.327 (1.822) &
  2 &
  3.890 (0.063) &
  \textbf{1} &
  \textbf{0.487 (0.009)} &
  \multicolumn{2}{c|}{} &
  \multicolumn{2}{c|}{} &
  \multicolumn{2}{c|}{} &
  \multicolumn{2}{c|}{} &
  \multicolumn{2}{c}{} \\
 &
  D-SOGA &
  2 &
  3.488 (0.079) &
  2 &
  6.541 (1.967) &
  2 &
  3.898 (0.150) &
  2 &
  0.487 (0.01) &
  \multicolumn{2}{c|}{} &
  \multicolumn{2}{c|}{} &
  \multicolumn{2}{c|}{} &
  \multicolumn{2}{c|}{} &
  \multicolumn{2}{c}{} \\
 &
  LiDOS &
  3 &
  3.501 (0.154) &
  2 &
  6.567 (2.086) &
  2 &
  3.897 (0.148) &
  2 &
  0.489 (0.01) &
  \multicolumn{2}{c|}{} &
  \multicolumn{2}{c|}{} &
  \multicolumn{2}{c|}{} &
  \multicolumn{2}{c|}{} &
  \multicolumn{2}{c}{} \\
\multirow{-5}{*}{\textbf{W12}} &
  DLiSA &
  2 &
  3.477 (0.065) &
  \cellcolor{teal!30}\textbf{1} &
  \cellcolor{teal!30}\textbf{5.341 (1.318)} &
  \cellcolor{teal!30}\textbf{1} &
  \cellcolor{teal!30}\textbf{3.878 (0.009)} &
  \cellcolor{teal!30}\textbf{1} &
  \cellcolor{teal!30}\textbf{0.487 (0.007)} &
  \multicolumn{2}{c|}{\multirow{-5}{*}{N/A}} &
  \multicolumn{2}{c|}{\multirow{-5}{*}{N/A}} &
  \multicolumn{2}{c|}{\multirow{-5}{*}{N/A}} &
  \multicolumn{2}{c|}{\multirow{-5}{*}{N/A}} &
  \multicolumn{2}{c}{\multirow{-5}{*}{N/A}} \\ \hline
 &
  FEMOSAA &
  3 &
  2.546 (0.024) &
  2 &
  3.621 (0.990) &
  \multicolumn{2}{c|}{} &
  \multicolumn{2}{c|}{} &
  \multicolumn{2}{c|}{} &
  \multicolumn{2}{c|}{} &
  \multicolumn{2}{c|}{} &
  \multicolumn{2}{c|}{} &
  \multicolumn{2}{c}{} \\
 &
  Seed-EA &
  2 &
  2.529 (0.018) &
  \textbf{1} &
  \textbf{3.440 (1.002)} &
  \multicolumn{2}{c|}{} &
  \multicolumn{2}{c|}{} &
  \multicolumn{2}{c|}{} &
  \multicolumn{2}{c|}{} &
  \multicolumn{2}{c|}{} &
  \multicolumn{2}{c|}{} &
  \multicolumn{2}{c}{} \\
 &
  D-SOGA &
  \cellcolor{red!20}\textbf{1} &
  \cellcolor{red!20}\textbf{2.527 (0.016)} &
  \textbf{1} &
  \textbf{3.554 (0.992)} &
  \multicolumn{2}{c|}{} &
  \multicolumn{2}{c|}{} &
  \multicolumn{2}{c|}{} &
  \multicolumn{2}{c|}{} &
  \multicolumn{2}{c|}{} &
  \multicolumn{2}{c|}{} &
  \multicolumn{2}{c}{} \\
 &
  LiDOS &
  3 &
  2.534 (0.019) &
  \textbf{1} &
  \textbf{3.612 (1.169)} &
  \multicolumn{2}{c|}{} &
  \multicolumn{2}{c|}{} &
  \multicolumn{2}{c|}{} &
  \multicolumn{2}{c|}{} &
  \multicolumn{2}{c|}{} &
  \multicolumn{2}{c|}{} &
  \multicolumn{2}{c}{} \\
\multirow{-5}{*}{\textbf{W13}} &
  DLiSA &
  3 &
  2.530 (0.018) &
  \cellcolor{teal!30}\textbf{1} &
  \cellcolor{teal!30}\textbf{2.939 (0.721)} &
  \multicolumn{2}{c|}{\multirow{-5}{*}{N/A}} &
  \multicolumn{2}{c|}{\multirow{-5}{*}{N/A}} &
  \multicolumn{2}{c|}{\multirow{-5}{*}{N/A}} &
  \multicolumn{2}{c|}{\multirow{-5}{*}{N/A}} &
  \multicolumn{2}{c|}{\multirow{-5}{*}{N/A}} &
  \multicolumn{2}{c|}{\multirow{-5}{*}{N/A}} &
  \multicolumn{2}{c}{\multirow{-5}{*}{N/A}} \\ \hline
 &
  FEMOSAA &
  \multicolumn{2}{c|}{\textbf{0}} &
  \multicolumn{2}{c|}{\textbf{4}} &
  \multicolumn{2}{c|}{\textbf{6}} &
  \multicolumn{2}{c|}{\textbf{0}} &
  \multicolumn{2}{c|}{\textbf{0}} &
  \multicolumn{2}{c|}{\textbf{1}} &
  \multicolumn{2}{c|}{\textbf{0}} &
  \multicolumn{2}{c|}{\textbf{0}} &
  \multicolumn{2}{c}{\textbf{0}} \\
 &
  Seed-EA &
  \multicolumn{2}{c|}{\textbf{5}} &
  \multicolumn{2}{c|}{\textbf{9}} &
  \multicolumn{2}{c|}{\textbf{2}} &
  \multicolumn{2}{c|}{\textbf{6}} &
  \multicolumn{2}{c|}{\textbf{5}} &
  \multicolumn{2}{c|}{\textbf{0}} &
  \multicolumn{2}{c|}{\textbf{2}} &
  \multicolumn{2}{c|}{\textbf{4}} &
  \multicolumn{2}{c}{\textbf{0}} \\
 &
  D-SOGA &
  \multicolumn{2}{c|}{\textbf{3}} &
  \multicolumn{2}{c|}{\textbf{9}} &
  \multicolumn{2}{c|}{\textbf{6}} &
  \multicolumn{2}{c|}{\textbf{4}} &
  \multicolumn{2}{c|}{\textbf{0}} &
  \multicolumn{2}{c|}{\textbf{0}} &
  \multicolumn{2}{c|}{\textbf{2}} &
  \multicolumn{2}{c|}{\textbf{5}} &
  \multicolumn{2}{c}{\textbf{0}} \\
 &
  LiDOS &
  \multicolumn{2}{c|}{\textbf{0}} &
  \multicolumn{2}{c|}{\textbf{4}} &
  \multicolumn{2}{c|}{\textbf{4}} &
  \multicolumn{2}{c|}{\textbf{0}} &
  \multicolumn{2}{c|}{\textbf{0}} &
  \multicolumn{2}{c|}{\textbf{0}} &
  \multicolumn{2}{c|}{\textbf{0}} &
  \multicolumn{2}{c|}{\textbf{2}} &
  \multicolumn{2}{c}{\textbf{0}} \\
\multirow{-5}{*}{\textbf{\begin{tabular}[c]{@{}c@{}}Summary \\ (\bm{$r$} = 1)\end{tabular}}} &
  DLiSA &
  \multicolumn{2}{c|}{\textbf{5}} &
  \multicolumn{2}{c|}{\textbf{13}} &
  \multicolumn{2}{c|}{\textbf{8}} &
  \multicolumn{2}{c|}{\textbf{6}} &
  \multicolumn{2}{c|}{\textbf{6}} &
  \multicolumn{2}{c|}{\textbf{8}} &
  \multicolumn{2}{c|}{\textbf{9}} &
  \multicolumn{2}{c|}{\textbf{8}} &
  \multicolumn{2}{c}{\textbf{6}} \\ \hline
\end{tabular}
}
	\end{adjustbox}
	\label{tb:RQ1_effectiveness_comparison}
\end{table*}

\subsubsection{Three or More Comparisons}
We leverage the Scott-Knott test \cite{mittas2012ranking} to compare multiple approaches. In a nutshell, it first ranks the approaches based on the mean performance scores and then iteratively partitions this ordered list into statistically distinct subgroups. These subgroups are determined by maximizing the inter-group mean square difference \(\Delta\) and their effect sizes. For example, for three approaches A, B, and C, the Scott-Knott test may yield two groups: \{A, B\} with rank 1 and \{C\} with rank 2, meaning that A and B are statistically similar but they are both significantly better than C.

	


\section{Results and Analysis}
\label{section:experimental_studies}


\subsection{RQ1: Effectiveness}

\subsubsection{Method} To answer \textbf{RQ1}, we compare \approach~with four state-of-the-art approaches discussed in Section \ref{subsection:compared_adaptation_approaches}. We aggregate and scrutinize the best-performing configurations from 100 independent runs (each with randomly ordered workloads) under every workload, across a total of 93 cases (9 systems and each with 6 to 13 workloads). We also use the Scott-Knott test \cite{mittas2012ranking}  for our analysis. All other settings are the same as discussed in Section \ref{subsection:component_and_parameter_settings}.


\subsubsection{Result} The experimental results are summarized in Table~\ref{sec:rq1}. As can be seen, overall, \approach~demonstrates superior performance, ranking first in 69 out of 93 cases, while \texttt{FEMOSAA}, \texttt{Seed-EA}, \texttt{D-SOGA}, and \texttt{LiDOS} achieve the best ranks in 11, 33, 29, and 10 out of 93 cases, respectively. Notably, within the 69 cases where \approach~achieved first rank, it also realized the best performance values in 65 cases, highlighting the efficacy and robustness of \approach~in self-adaptation. In particular, \approach~achieves up to 2.29$\times$ enhancement compared with its counterparts (W8 of the \textsc{kanzi}).

The above efficacy of \approach~lies in its knowledge distillation for seeding, tailored to the characteristics of changing workloads in configurable systems. This empowers \approach~to judiciously and dynamically distill the most useful configurations and ignore the misleading ones to enhance planning or engage a conservative stance in case the past configurations are generally useless. Other approaches, in contrast, either ignore the valuable past knowledge or leverage it without catering to the noise, due to the stationary setting and the static knowledge exploitation strategy.

However, there are also some edge cases where other approaches are competitive to \approach. For instance, in \textsc{lrzip}, \textsc{dconvert}, and \textsc{batlik} systems, the relatively high similarity across different workloads suggests that consistently effective configuration in one workload tends to perform well in others. This consistency favors the \texttt{Seed-EA} approach, which simply seeds all the configurations preserved in preceded planning without specific responses to workload changes. An interesting observation arises within \textsc{z3} system, in which 
\texttt{D-SOGA} exhibits relatively superior performance. This could be attributed to a possible moderate similarity across workloads, which allows the combination of historical insights and random configurations in \texttt{D-SOGA} to thrive.

Based on the above analysis, we can conclude that:

\keystate{
\textit{\textbf{RQ1:} \approach~is effective as it is generally ranked better (in the statistical sense) than state-of-the-art in 74\% cases (69 out of 93) with significant performance improvements of up to 2.29$\times$.}
}


\begin{table}[t!]
	\centering
	\caption{Comparing resource efficiency of \approach~with respect to the state-of-the-art approaches. Detailed results can be found in our repository: \begin{small}\revision{\texttt{\url{https://github.com/ideas-labo/dlisa}}}\end{small}.}
	\label{tb:efficiency}
        \begin{adjustbox}{width=\columnwidth,center}
			\small
\footnotesize
\setlength{\tabcolsep}{2pt}

	\begin{tabular}{l|cccc|l|cccc|l|cccc|l|cccc}
		\toprule
		\multirow{4}{*}{\textbf{System}} & \multicolumn{4}{c|}{\textbf{\texttt{FEMOSAA}}}                                                                                                            &  & \multicolumn{4}{c|}{\textbf{\texttt{Seed-EA}}}                                                                                                            &  & \multicolumn{4}{c|}{\textbf{\texttt{D-SOGA}}}                                                                                                             &  & \multicolumn{4}{c}{\textbf{\texttt{LiDOS}}}                                                                                                               \\ \cline{2-5} \cline{7-10} \cline{12-15} \cline{17-20} 
		& \multicolumn{1}{c|}{\rotatebox[origin=c]{90}{$s$ \textgreater 1}} & \multicolumn{1}{c|}{\rotatebox[origin=c]{90}{$s$ = 1}} & \multicolumn{1}{c|}{\rotatebox[origin=c]{90}{0 \textless{} $s$ \textless 1}} & \rotatebox[origin=c]{90}{$s$ = N/A} &  & \multicolumn{1}{c|}{\rotatebox[origin=c]{90}{$s$ \textgreater 1}} & \multicolumn{1}{c|}{\rotatebox[origin=c]{90}{$s$ = 1}} & \multicolumn{1}{c|}{\rotatebox[origin=c]{90}{0 \textless{} $s$ \textless 1}} & \rotatebox[origin=c]{90}{$s$ = N/A} &  & \multicolumn{1}{c|}{\rotatebox[origin=c]{90}{$s$ \textgreater 1}} & \multicolumn{1}{c|}{\rotatebox[origin=c]{90}{$s$ = 1}} & \multicolumn{1}{c|}{\rotatebox[origin=c]{90}{0 \textless{} $s$ \textless 1}} & \rotatebox[origin=c]{90}{$s$ = N/A} &  & \multicolumn{1}{c|}{\rotatebox[origin=c]{90}{$s$ \textgreater 1}} & \multicolumn{1}{c|}{\rotatebox[origin=c]{90}{$s$ = 1}} & \multicolumn{1}{c|}{\rotatebox[origin=c]{90}{0 \textless{} $s$ \textless 1}} & \rotatebox[origin=c]{90}{$s$ = N/A} \\ \cline{1-5} \cline{7-10} \cline{12-15} \cline{17-20} 
		\textsc{lrzip}                   & \multicolumn{1}{c|}{\textbf{13}}                 & \multicolumn{1}{c|}{0}       & \multicolumn{1}{c|}{0}                             & 0         &  & \multicolumn{1}{c|}{2}                  & \multicolumn{1}{c|}{2}       & \multicolumn{1}{c|}{1}                             & \textbf{8}         &  & \multicolumn{1}{c|}{\textbf{6}}                  & \multicolumn{1}{c|}{2}       & \multicolumn{1}{c|}{0}                             & 5         &  & \multicolumn{1}{c|}{\textbf{10}}                 & \multicolumn{1}{c|}{3}       & \multicolumn{1}{c|}{0}                             & 0         \\
		\textsc{xz}                      & \multicolumn{1}{c|}{\textbf{13}}                 & \multicolumn{1}{c|}{0}       & \multicolumn{1}{c|}{0}                             & 0         &  & \multicolumn{1}{c|}{\textbf{13}}                 & \multicolumn{1}{c|}{0}       & \multicolumn{1}{c|}{0}                             & 0         &  & \multicolumn{1}{c|}{\textbf{13}}                 & \multicolumn{1}{c|}{0}       & \multicolumn{1}{c|}{0}                             & 0         &  & \multicolumn{1}{c|}{\textbf{13}}                 & \multicolumn{1}{c|}{0}       & \multicolumn{1}{c|}{0}                             & 0         \\
		\textsc{z3}                      & \multicolumn{1}{c|}{\textbf{8}}                  & \multicolumn{1}{c|}{1}       & \multicolumn{1}{c|}{0}                             & 3         &  & \multicolumn{1}{c|}{\textbf{8}}                  & \multicolumn{1}{c|}{3}       & \multicolumn{1}{c|}{0}                             & 1         &  & \multicolumn{1}{c|}{\textbf{6}}                  & \multicolumn{1}{c|}{2}       & \multicolumn{1}{c|}{0}                             & 4         &  & \multicolumn{1}{c|}{\textbf{7}}                  & \multicolumn{1}{c|}{3}       & \multicolumn{1}{c|}{0}                             & 2         \\
		\textsc{dconvert}                & \multicolumn{1}{c|}{\textbf{12}}                 & \multicolumn{1}{c|}{0}       & \multicolumn{1}{c|}{0}                             & 0         &  & \multicolumn{1}{c|}{1}                  & \multicolumn{1}{c|}{\textbf{6}}       & \multicolumn{1}{c|}{0}                             & 5         &  & \multicolumn{1}{c|}{1}                  & \multicolumn{1}{c|}{\textbf{6}}       & \multicolumn{1}{c|}{0}                             & 5         &  & \multicolumn{1}{c|}{\textbf{11}}                 & \multicolumn{1}{c|}{0}       & \multicolumn{1}{c|}{0}                             & 1         \\
		\textsc{batlik}                  & \multicolumn{1}{c|}{\textbf{11}}                 & \multicolumn{1}{c|}{0}       & \multicolumn{1}{c|}{0}                             & 0         &  & \multicolumn{1}{c|}{2}                  & \multicolumn{1}{c|}{4}       & \multicolumn{1}{c|}{0}                             & \textbf{5}         &  & \multicolumn{1}{c|}{3}                  & \multicolumn{1}{c|}{\textbf{6}}       & \multicolumn{1}{c|}{0}                             & 2         &  & \multicolumn{1}{c|}{\textbf{6}}                  & \multicolumn{1}{c|}{4}       & \multicolumn{1}{c|}{0}                             & 1         \\
		\textsc{kanzi}                   & \multicolumn{1}{c|}{\textbf{8}}                  & \multicolumn{1}{c|}{0}       & \multicolumn{1}{c|}{0}                             & 1         &  & \multicolumn{1}{c|}{\textbf{9}}                  & \multicolumn{1}{c|}{0}       & \multicolumn{1}{c|}{0}                             & 0         &  & \multicolumn{1}{c|}{\textbf{9}}                  & \multicolumn{1}{c|}{0}       & \multicolumn{1}{c|}{0}                             & 0         &  & \multicolumn{1}{c|}{\textbf{9}}                  & \multicolumn{1}{c|}{0}       & \multicolumn{1}{c|}{0}                             & 0         \\
		\textsc{x264}                    & \multicolumn{1}{c|}{\textbf{9}}                  & \multicolumn{1}{c|}{0}       & \multicolumn{1}{c|}{0}                             & 0         &  & \multicolumn{1}{c|}{\textbf{9}}                  & \multicolumn{1}{c|}{0}       & \multicolumn{1}{c|}{0}                             & 0         &  & \multicolumn{1}{c|}{\textbf{9}}                  & \multicolumn{1}{c|}{0}       & \multicolumn{1}{c|}{0}                             & 0         &  & \multicolumn{1}{c|}{\textbf{9}}                  & \multicolumn{1}{c|}{0}       & \multicolumn{1}{c|}{0}                             & 0         \\
		\textsc{h2}                      & \multicolumn{1}{c|}{\textbf{8}}                  & \multicolumn{1}{c|}{0}       & \multicolumn{1}{c|}{0}                             & 0         &  & \multicolumn{1}{c|}{\textbf{7}}                  & \multicolumn{1}{c|}{1}       & \multicolumn{1}{c|}{0}                             & 0         &  & \multicolumn{1}{c|}{\textbf{5}}                  & \multicolumn{1}{c|}{2}       & \multicolumn{1}{c|}{0}                             & 1         &  & \multicolumn{1}{c|}{\textbf{8}}                  & \multicolumn{1}{c|}{0}       & \multicolumn{1}{c|}{0}                             & 0         \\
		\textsc{jump3r}                  & \multicolumn{1}{c|}{\textbf{6}}                  & \multicolumn{1}{c|}{0}       & \multicolumn{1}{c|}{0}                             & 0         &  & \multicolumn{1}{c|}{\textbf{6}}                  & \multicolumn{1}{c|}{0}       & \multicolumn{1}{c|}{0}                             & 0         &  & \multicolumn{1}{c|}{\textbf{6}}                  & \multicolumn{1}{c|}{0}       & \multicolumn{1}{c|}{0}                             & 0         &  & \multicolumn{1}{c|}{\textbf{6}}                  & \multicolumn{1}{c|}{0}       & \multicolumn{1}{c|}{0}                             & 0         \\ \cline{1-5} \cline{7-10} \cline{12-15} \cline{17-20} 
		\textbf{Total}                   & \multicolumn{1}{c|}{\textbf{88}}                 & \multicolumn{1}{c|}{1}       & \multicolumn{1}{c|}{0}                             & 4         &  & \multicolumn{1}{c|}{\textbf{57}}                 & \multicolumn{1}{c|}{16}      & \multicolumn{1}{c|}{1}                             & 19        &  & \multicolumn{1}{c|}{\textbf{58}}                 & \multicolumn{1}{c|}{18}      & \multicolumn{1}{c|}{0}                             & 17        &  & \multicolumn{1}{c|}{\textbf{79}}                 & \multicolumn{1}{c|}{10}       & \multicolumn{1}{c|}{0}                             & 4         \\ \cline{1-5} \cline{7-10} \cline{12-15} \cline{17-20} 
		\textbf{Range of $s$}              & \multicolumn{4}{c|}{$s \in [1, 2.16]$}                                                                                                  &  & \multicolumn{4}{c|}{$s \in (0, 2.22]$}                                                                                                  &  & \multicolumn{4}{c|}{$s \in [1, 2.05]$}                                                                                                  &  & \multicolumn{4}{c}{$s \in [1, 2.05]$}                                                                                                   \\ \bottomrule
	\end{tabular}
	\end{adjustbox}

\end{table}

\subsection{RQ2: Efficiency}

\subsubsection{Method} To evaluate the resource efficiency in \textbf{RQ2}, for each case out of the 93, we employ the following procedure:

\begin{itemize}
	\item A baseline, $b$, is identified for each counterpart approach, representing the smallest number of measurements necessary for it to reach its best performance, \revision{denoted as $T$}, averaging over 100 runs.
	
	\item For \approach, find the smallest number of measurements, denote as $m$, at which the average result of the performance (over 100 runs) is equivalent to or better than $T$.
	
	\item The speedup of \approach~over a counterpart is reported as $s = \frac{b}{m}$, which is a common metric used in \cite{gao2021resource,chen2021mmo}.
\end{itemize}

If \approach~is efficient, then we expect $s>1$; $0<s<1$ and $s=1$ means \approach~has worse efficiency and they are equally efficient, respectively. We use $s=$ N/A to denote the case where \approach~cannot achieve the $T$ reached by its counterpart. All other settings are the same as \textbf{RQ1}.



\subsubsection{Result}  


The results are depicted in Table \ref{tb:efficiency}, clearly, \approach~consistently outperforms or equalling its counterparts in the majority of cases. Specifically, compared with \texttt{FEMOSAA}, \approach~attained superior speedup in 88 cases (up to 2.16$\times$) with equal efficiency in 1 case. When contrasted to \texttt{Seed-EA}, \approach~excelled in 57 cases with a maximum of 2.22$\times$ speedup and matched in 16. For the comparisons with \texttt{D-SOGA} and \texttt{LiDOS}, \approach~maintained a remarkable speedup within 58 and 79 (up to 2.05$\times$) out of 93 cases, respectively. These observations illustrate \approach's ability to deliver robust performance in utilizing resources for self-adapting configurable systems. Therefore, we say:


\keystate{
\textit{\textbf{RQ2:} \approach~is considerably more efficient than state-of-the-art approaches in the majority of the cases, achieving up to $2.22\times$ speedup. }
\vspace{-0.08cm}
}


\subsection{RQ3: Ablation Analysis}

\subsubsection{Method} 

To understand which parts in the knowledge distillation of \approach~work, in \textbf{RQ3}, we design two variants to compare with the original \approach~over the 93 cases:

\begin{itemize}
    \item \textbf{\approach-\texttt{I}:} We replace the weighted configuration seeding with a random seeding of preserved past configurations.
    \item \textbf{\approach-\texttt{II}:} We disable the ranked workload similarity analysis but randomly trigger the seeding of planning. 
\end{itemize}

Since there are only pairwise comparisons, we leverage the Wilcoxon rank-sum test and $\hat{A}_{12}$ effect size across 100 runs.


\begin{table}[t]
	\caption{Comparing \approach~against its two variants over 100 runs; “+”, “=”, and “$\mathbf{-}$” respectively indicate \approach~performing significantly better than, similarly to, or worse than the variants. Detailed results can be found in our repository: \begin{small}\revision{\texttt{\url{https://github.com/ideas-labo/dlisa}}}\end{small}.}
	\label{tb:ablation_summary}
	\centering
	
\begin{tabular}{lcc}
	\toprule
	\textbf{System} & \textbf{\approach~vs \approach-\texttt{I}} & \textbf{\approach~vs \approach-\texttt{II}} \\ \midrule
	\textsc{lrzip}           & 5+/8=/0$\mathbf{-}$           & 5+/8=/0$\mathbf{-}$              \\
	\textsc{xz}              & 13+/0=/0$\mathbf{-}$            & 5+/8=/0$\mathbf{-}$              \\
	\textsc{z3}              & 0+/12=/0$\mathbf{-}$            & 3+/8=/1$\mathbf{-}$              \\
	\textsc{dconvert}        & 2+/10=/0$\mathbf{-}$           & 5+/7=/0$\mathbf{-}$             \\
	\textsc{batlik}          & 5+/6=/0$\mathbf{-}$             & 10+/1=/0$\mathbf{-}$             \\
	\textsc{kanzi}           & 8+/1=/0$\mathbf{-}$             & 6+/3=/0$\mathbf{-}$              \\
	\textsc{x264}            & 9+/0=/0$\mathbf{-}$             & 0+/9=/0$\mathbf{-}$             \\
	\textsc{h2}              & 2+/6=/0$\mathbf{-}$            & 2+/6=/0$\mathbf{-}$              \\
	\textsc{jump3r}          & 6+/0=/0$\mathbf{-}$             & 3+/3=/0$\mathbf{-}$              \\ \midrule
	\textbf{Total}  & \textbf{50+/43=/0$\mathbf{-}$ } & \textbf{39+/53=/1$\mathbf{-}$ }  \\ \bottomrule
\end{tabular}

\end{table}

\subsubsection{Result} The results are summarized in Table \ref{tb:ablation_summary}. Clearly, we see that \approach~exhibits a remarkable improvement over its variants from the 93 cases: \approach~wins {\approach-\texttt{I}} in 50 cases with 43 ties, reflecting the effectiveness of weighted configuration seeding. Against {\approach-\texttt{II}}, \approach~wins in 39 cases; draws in 53 cases; and loses only in one case, indicating the usefulness of the workload similarity analysis. These results prove the benefit of seeding only when needed and the positive implication of considering the most useful configurations while excluding the misleading ones---all of which are specifically designed in our knowledge distillation according to the key characteristic of configurable systems under changing workloads discussed in Section \ref{subsection:motivation_and_challenges}.




In light of these observations, we can conclude that:


\keystate{
\textit{\textbf{RQ3:} Each individual parts in the knowledge distillation in \approach~contribute significantly to its superiority.}
}

\subsection{RQ4: Sensitivity to $\alpha$} 
\label{subsection:sensitivity_to_alpha}
\subsubsection{Method} 

The parameter $\alpha$ determines the likelihood of triggering seeding, in \textbf{RQ4}, we examine the sensitivity of \approach~to $\alpha$ by comparing the cases of $\alpha \in \{0,0.1,...,0.9\}$. Again, we use the Scott-Knott to compare the results against different $\alpha$ values over 100 runs for all cases.

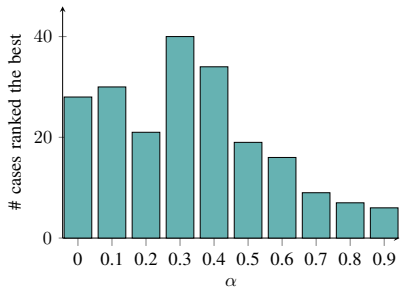
\begin{figure}[t!]
\centering
\begin{tikzpicture}
\begin{axis}[
height=6cm, width=8cm,
grid style = {dashed},
      axis x line  = bottom,
    axis y line  = left  ,
legend pos=outer north east,
ymin=0, ymax=46,
enlarge x limits=0.05,
ybar=0pt,
bar width=15pt,
xlabel={$\alpha$},
xtick={1,2,3,4,5,6,7,8,9,10,11,12,13,14,15,16,17,18,19,20,21,22},
xticklabels={0,0.1,0.2,0.3,0.4,0.5,0.6,0.7,0.8,0.9},
ylabel={\textcolor{black}{\# cases ranked the best}},
y axis line style={black},
yticklabel style={black},
xticklabel style={rotate=0},
legend columns=1,
legend cell align={left},
  legend style={nodes={scale=1, transform shape},draw=none,fill=none,at={(0.7,0.95)}},
        extra y tick labels={},
        extra y tick style={grid=major,major grid style={thick,draw=black}}
]

\addplot[fill=teal!60,opacity=1,draw=black]
coordinates { 
(1,28)
(2,30)
(3,21)
(4,40)
(5,34)
(6,19)
(7,16)
(8,9)
(9,7)
(10,6)
};


\end{axis}

\end{tikzpicture}
\caption{The sensitivity of \approach~to different $\alpha$ values.}
\label{fig:alhpa}
\end{figure}


\subsubsection{Result} As illustrated in Figure~\ref{fig:alhpa}, \approach~exhibits obviously superior performance when the $\alpha$ is set to 0.3 in terms of Scott-Knott ranks. We see that neither too small nor too large $\alpha$ is optimal. This is because, in the former case, there would be too many unnecessary seeding, making it difficult to eliminate the misleading information even with the weighted configuration seeding (e.g., $\alpha=0$ means seeding constantly). In the latter case, it is simply due to the fact that it becomes rather difficult to trigger seeding, but instead rely merely on random initialization and hence waste the valuable accumulated knowledge. Clearly, the degradation of having larger $\alpha$ is more serious than setting it small, since not being able to seed is more influential than seeding misleading noises on the systems/workloads considered. Overall, we show that:



\keystate{
\textit{\textbf{RQ4:} Setting $\alpha$ to 0.3 yields the most effective performance for \approach, as it reaches a better balance between the benefit of seeding and seeding misleading information.}
}

        \section{Discussion}
\label{section:discussion}

\subsection{How Workload Similarity Analysis Helps?}

To understand why the ranked workload similarity analysis can help, Figure~\ref{fig:diss}a shows the changing similarity scores on two exampled orders of the time-varying workloads for \textsc{z3}. As can be seen, the similarity scores differ depending on the sequence of the emergent workloads---in some cases, they are higher than the threshold $\alpha=0.3$ while in some other cases, they are lower. Such a discrepancy reflects the likelihood of seeding being beneficial: in the former cases, the seeding is constantly triggered because configurations found via the planning are sufficiently similar; while in the latter cases, randomly initialized configurations are used instead as seeding would likely be more harmful due to the misleading information caused by rather different landscapes between the workloads. In this way, \approach~retains robust adaptability to diverse and changing workloads on configurable systems.

\subsection{Why Weighted Configuration Seeding Work?}

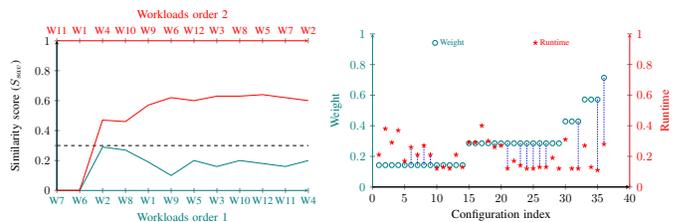
\begin{figure}[t!]
\centering
\subfloat[workload similarity (\textsc{z3})]
{\begin{tikzpicture}
  \begin{axis}[
height=6cm, width=9.0cm,
       ymin=0,
    ymax=1,
    ylabel={\textcolor{black}{Similarity score ($S_{sav}$)}},
    xlabel={\textcolor{teal}{Workloads order 1}},
    xtick/.style={teal},
    ytick/.style={black},
    yticklabel style={black},
xticklabel style={teal},
x axis line style={teal},
    xtick={0,1,2,3,4,5,6,7,8,9,10,11,12,13,14,15},
    symbolic x coords={W7,W6,W2,W8,W1,W9,W5,W3,W10,W12,W11,W4},
    xtick distance=1.0, 
      axis x line  = bottom,
    axis y line  = left  ,
    extra y ticks = 0.3,
     extra y tick labels={},
     xticklabel style = {font=\small},
        extra y tick style={grid=major,major grid style={dashed,draw=black}},
    legend cell align={left},
    legend style={draw=none,fill=none,at={(0.95,0.95)}},             
  ] 


 \addplot [no marks,teal] plot coordinates {
(W7,0.0)
(W6,0.0)
(W2,0.29)
(W8,0.27)
(W1,0.19)
(W9,0.1)
(W5,0.2)
(W3,0.16)
(W10,0.2)
(W12,0.18)
(W11,0.16)
(W4,0.2)

 };


   
    
  \end{axis}

  \begin{axis}[
height=6cm, width=9.0cm,
    ymin=0,
    ymax=1,
    xlabel={\textcolor{red}{Workloads order 2}},
    xtick/.style={red},
xticklabel style={red},
x axis line style={red},
y axis line style={black},
      xtick={0,1,...,12},
     symbolic x coords={W11,W1,W4,W10,W9,W6,W12,W3,W8,W5,W7,W2},  axis x line  = top,
         xtick distance=1.0,
    axis y line  = left  ,
    ytick=\empty,
    legend cell align={left},
      xticklabel style = {font=\small},
    legend style={draw=none,fill=none,at={(0.95,0.95)}},             
  ]

 \addplot [no marks,red] plot coordinates {
(W11,0)
(W1,0)
(W4,0.47)
(W10,0.46)
(W9,0.57)
(W6,0.62)
(W12,0.6)
(W3,0.63)
(W8,0.63)
(W5,0.64)
(W7,0.62)
(W2,0.6)
 };


    
  \end{axis}

\end{tikzpicture}}
~\hfill
\subfloat[configuration weights (\textsc{kanzi})]
{	
	\begin{tikzpicture}
		\begin{axis}[
			height=6cm, width=9.0cm,
			xlabel={\textcolor{black}{Configuration index}},
			ylabel={\textcolor{teal}{Weight}},
			ytick/.style={teal},
			yticklabel style={teal},
			y axis line style={teal},
			axis x line  = bottom,
			axis y line  = left,
			legend cell align={left},
			legend style={draw=none,fill=none,at={(0.95,0.95)}}, 
			ymin=0, ymax=1,
			xmin=0, xmax=40,
			legend style={
				at={(0.3,1.0)}, 
				anchor=north,
				legend columns=-1, 
				font=\fontsize{7}{11}\selectfont
			}
			]


			\addplot [mark=o,teal,only marks] plot coordinates {
				(1,0.14285714285714285)
				(2,0.14285714285714285)
				(3,0.14285714285714285)
				(4,0.14285714285714285)
				(5,0.14285714285714285)
				(6,0.14285714285714285)
				(7,0.14285714285714285)
				(8,0.14285714285714285)
				(9,0.14285714285714285)
				(10,0.14285714285714285)
				(11,0.14285714285714285)
				(12,0.14285714285714285)
				(13,0.14285714285714285)
				(14,0.14285714285714285)
				(15,0.2857142857142857)
				(16,0.2857142857142857)
				(17,0.2857142857142857)
				(18,0.2857142857142857)
				(19,0.2857142857142857)
				(20,0.2857142857142857)
				(21,0.2857142857142857)
				(22,0.2857142857142857)
				(23,0.2857142857142857)
				(24,0.2857142857142857)
				(25,0.2857142857142857)
				(26,0.2857142857142857)
				(27,0.2857142857142857)
				(28,0.2857142857142857)
				(29,0.2857142857142857)
				(30,0.42857142857142855)
				(31,0.42857142857142855)
				(32,0.42857142857142855)
				(33,0.5714285714285714)
				(34,0.5714285714285714)
				(35,0.5714285714285714)
				(36,0.7142857142857143)
			};
			\addlegendentry{\textcolor{teal}{Weight}}

			
			
			
		\end{axis}

		\begin{axis}[
			height=6cm, width=9.0cm,
			ylabel={\textcolor{red}{Runtime}},
			ytick/.style={red},
			yticklabel style={red},
			y axis line style={red},
			axis x line  = bottom,
			axis y line  = right  ,
			xtick=\empty,
			legend cell align={left},
			legend style={draw=none,fill=none,at={(0.95,0.95)}},  
			ymin=0, ymax=1,
			xmin=0, xmax=40,
			legend style={
				at={(0.7,1.0)}, 
				anchor=north,
				font=\fontsize{7}{11}\selectfont
			}
			]

			\addplot [mark=star,red,only marks] plot coordinates {
				(1,0.21)
				(2,0.38)
				(3,0.29)
				(4,0.37)
				(5,0.17)
				(6,0.26)
				(7,0.21)
				(8,0.27)
				(9,0.21)
				(10,0.12)
				(11,0.13)
				(12,0.12)
				(13,0.21)
				(14,0.13)
				(15,0.29)
				(16,0.29)
				(17,0.4)
				(18,0.3)
				(19,0.26)
				(20,0.27)
				(21,0.12)
				(22,0.17)
				(23,0.14)
				(24,0.12)
				(25,0.12)
				(26,0.13)
				(27,0.13)
				(28,0.19)
				(29,0.12)
				(30,0.31)
				(31,0.12)
				(32,0.12)
				(33,0.27)
				(34,0.13)
				(35,0.11)
				(36,0.28)
				
			};
			\addlegendentry{\textcolor{red}{Runtime}}
	
			\draw [color=blue, dashed, line width = 0.2 mm, dash pattern=on 1pt off 1pt] (axis cs:27,0.2857142857142857) -- (axis cs:27,0.13);
			\draw [color=blue, dashed, line width = 0.2 mm, dash pattern=on 1pt off 1pt] (axis cs:6,0.14285714285714285) -- (axis cs:6,0.26);
			\draw [color=blue, dashed, line width = 0.2 mm, dash pattern=on 1pt off 1pt] (axis cs:19,0.2857142857142857) -- (axis cs:19,0.26);
			\draw [color=blue, dashed, line width = 0.2 mm, dash pattern=on 1pt off 1pt] (axis cs:21,0.2857142857142857) -- (axis cs:21,0.12);
			\draw [color=blue, dashed, line width = 0.2 mm, dash pattern=on 1pt off 1pt] (axis cs:26,0.2857142857142857) -- (axis cs:26,0.13);
			\draw [color=blue, dashed, line width = 0.2 mm, dash pattern=on 1pt off 1pt] (axis cs:32,0.42857142857142855) -- (axis cs:32,0.12);
			\draw [color=blue, dashed, line width = 0.2 mm, dash pattern=on 1pt off 1pt] (axis cs:9,0.14285714285714285) -- (axis cs:9,0.21);
			\draw [color=blue, dashed, line width = 0.2 mm] (axis cs:10,0.14285714285714285) -- (axis cs:10,0.12);
			\draw [color=blue, dashed, line width = 0.2 mm, dash pattern=on 1pt off 1pt] (axis cs:23,0.2857142857142857) -- (axis cs:23,0.14);
			\draw [color=blue, dashed, line width = 0.2 mm, dash pattern=on 1pt off 1pt] (axis cs:24,0.2857142857142857) -- (axis cs:24,0.12);
			\draw [color=blue, dashed, line width = 0.2 mm, dash pattern=on 1pt off 1pt] (axis cs:35,0.5714285714285714) -- (axis cs:35,0.11);
			\draw [color=blue, dashed, line width = 0.2 mm, dash pattern=on 1pt off 1pt] (axis cs:25,0.2857142857142857) -- (axis cs:25,0.12);
			\draw [color=blue, dashed, line width = 0.2 mm, dash pattern=on 1pt off 1pt] (axis cs:7,0.14285714285714285) -- (axis cs:7,0.21);
			\draw [color=blue, dashed, line width = 0.2 mm] (axis cs:20,0.2857142857142857) -- (axis cs:20,0.27);
			\draw [color=blue, dashed, line width = 0.2 mm, dash pattern=on 1pt off 1pt] (axis cs:16,0.2857142857142857) -- (axis cs:16,0.29);
			\draw [color=blue, dashed, line width = 0.2 mm, dash pattern=on 1pt off 1pt] (axis cs:8,0.14285714285714285) -- (axis cs:8,0.27);
			\draw [color=blue, dashed, line width = 0.2 mm, dash pattern=on 1pt off 1pt] (axis cs:36,0.7142857142857143) -- (axis cs:36,0.28);
			
			\draw [color=blue, dashed, line width = 0.2 mm, dash pattern=on 1pt off 1pt] (axis cs:5,0.14285714285714285) -- (axis cs:5,0.17);
			\draw [color=blue, dashed, line width = 0.2 mm] (axis cs:11,0.14285714285714285) -- (axis cs:11,0.13);
			\draw [color=blue, dashed, line width = 0.2 mm, dash pattern=on 1pt off 1pt] (axis cs:15,0.2857142857142857) -- (axis cs:15,0.29);

			\addplot [draw=none] coordinates {}; 
			\addlegendentry{\textcolor{blue}{Connection}} 

			

			
		\end{axis}

	\end{tikzpicture}
	}

\caption{Examples illustrating the workload similarity and seeded configurations with respect to the weights. In (b), the dotted lines highlight the selected ones for seeding.}
\label{fig:diss}
\end{figure}

To demonstrate how configurations are weighted for knowledge distillation, Figure~\ref{fig:diss}b visualizes the seeds extraction process for self-adaptation planning under workload \texttt{artificl} on the \textsc{kanzi}. The weights of distilled configurations selected for seeding and their performance under the current workload are connected by dashed lines, in which we see that configurations with higher weights are often selected, and they generally yield excellent performance (i.e., smaller runtime) than most of the remaining ones. Since we select the configurations stochastically based on the weights, we see that a small set of those with lower weights is also selected. Those lower weighted configurations, albeit slightly worse than the others on performance, help to prevent the selection of too many similar configurations for seeding. The above is what makes the weighted configuration effective in \approach.



\subsection{What Are the Implications of \approach?}
\revision{Lifelong self-adaptation (or self-evolving systems), as highlighted in prior work~\cite{chen2022lifelong,DBLP:journals/jid/WeynsBVYB22}, is an emerging paradigm for ensuring that systems can evolve autonomously under unanticipated changes. This study works along this direction from the perspective of seeding under changing workloads. It shows that the evolutionary planning that runs continuously for self-adaptation is beneficial. We have demonstrated the effectiveness of the two key contributions designed in \approach---\textit{ranked workload similarity analysis} and \textit{weighted configuration seeding}---in achieving lifelong self-adaptation for configurable systems: \approach~considerably enhances system performance by identifying and leveraging useful historical knowledge while alleviating the impact of misleading information. These results hold substantial implications for the field of Software Engineering, as they support the development of more resilient and dependable software systems that make use of existing useful knowledge while filtering out useless knowledge at varying workloads. As such, we anticipate that our results will further advance the existing research in engineering self-adaptive configurable systems.}

	
	\section{Threats to Validity}
\label{section:threats_to_validity}

Our investigation acknowledges the potential threats to \textbf{internal validity} associated with the parameter $\alpha$, which we have set to 0.3. This choice is grounded in empirical evidence from our experiments in \textbf{RQ4}, where $\alpha = 0.3$ is a ``rule-of-thumb'' that yields generally favorable outcomes. \revision{We admit that the best value for $\alpha$ may differ on a system-by-system basis and that exploring different settings for $\alpha$ might enhance the robustness of our results.} For all experiments, we also use statistical tests and effect sizes to mitigate this threat.



Threats regarding \textbf{external validity} may arise from the specific configurable systems and workloads selected for our study. To mitigate potential biases, we included nine systems in our study. These systems span different domains, scales, and performance objectives, and we tested them across 93 unique workloads, following benchmarks set by prior studies \cite{muhlbauer2023analyzing}. This diverse selection aims to enhance the generalizability of our results, though expanding the range of systems examined could further deepen our insights.
	
	\section{Related Work}
\label{section:related-work}
Here, we discuss the related work in light of \approach.

\subsection{Stationary Adaptation for Configurable Systems}
Stationary adaptation in self-adaptive configurable systems has been the cornerstone of strategies aiming at tuning the configurations under time-varying workloads \cite{kumar2020datesso,kumar2018multi,li2020bilo,chen2015self}. Traditional approaches, such as those proposed by Chen \textit{et al.} \cite{chen2018femosaa} and similar frameworks \cite{elkhodary2010fusion,chen2020synergizing} primarily focus on planning from scratch when changes in workloads are detected or at regular intervals, neglecting the accumulation of valuable historical insights. While this simplifies the optimization process, it may lead to repetitive effort and ineffective planning.

In contrast, our proposed \approach~framework adopts a more holistic view that goes beyond the stationary paradigm. By harnessing the wealth of information available from past search experiences, \approach~aims to construct a lifelong planning trajectory for the system. In this way, \approach~sidesteps the inefficiencies of stationary planning, fostering a more intelligent optimization process that leverages historical information to facilitate future adaption planning.


\subsection{Dynamic Adaptation for Configurable Systems}
Dynamic adaptation is characterized by algorithms designed for planning continuously and adapting in real time. For example, Ramirez \textit{et al.} \cite{ramirez2009applying} propose \textsc{Plato}, a framework for adaptation planning using SOGA, which can automatically achieve adaptation planning by detecting the changes in fitness. Chen \textit{et al.}\cite{chen2018effects} and Kinneer \textit{et al.}\cite{kinneer2021information} also use the concept of seeding to expedite the planning. This process, inspired by the principles of natural evolution and state preservation, is aimed at seamlessly self-adapting the workload changes, thereby achieving ``lifelong optimization''. 


Although existing methods are dynamic in nature, they diverge from the true definition of dynamic optimization in the work \cite{nguyen2012evolutionary}. This deviation stems from the static knowledge exploitation strategy, where all (or randomly chosen) configurations from the most recent past workload are seeded for the current one, while those from earlier workloads are simply discarded. In contrast, \approach~introduces distilled knowledge seeding, a truly dynamic knowledge exploitation strategy, designed to navigate planning under time-varying workloads. Rather than naively assuming that only the configurations from the most recent past workload are useful, it proactively extracts the useful configurations for seeding from all past workloads while doing so only when it is necessary. 


\subsection{Control Theoretical Configuration Adaptation}
Control theory has been recognized as an effective solution for the planning of configurable systems \cite{filieri2015automated, maggio2017automated,shevtsov2016keep}. Among others, Maggio \textit{et al.} \cite{maggio2017automated} employ Kalman filters to refine and update the state values of the controller model, which are central to model predictive control schemes. Shevtsov and Weyns \cite{shevtsov2016keep} expand on this by incorporating the simplex optimization method, which targets global optima in system states, thereby enhancing the precision of control mechanisms. The application of control theory in self-adaptation planning is promising, however, it faces significant challenges due to the complex, non-linear dynamics of real systems that can only be prescribed with advanced domain knowledge \cite{zhu2009does}. 


\subsection{Configuration Performance Learning}
Configuration performance learning for configurable systems is a distinct research trajectory that focuses on modeling the correlation between configuration and performance~\cite{DBLP:journals/corr/abs-2403-03322}. Several methods have been used, such as support-vector machines \cite{yigitbasi2013towards}, decision trees \cite{nair2018faster}, neural network \cite{chen2013self,DBLP:journals/pacmse/Gong024,DBLP:conf/sigsoft/Gong023}, and ensemble learning \cite{chen2016self,DBLP:conf/sigsoft/Gong023,gongchentse2024}, with the goal of crafting a function that accurately encapsulates the correlation between adaptation options and the performance of the target system. In contrast, \approach~emphasizes optimization to self-adapting the changes of systems---a complementary aspect to performance learning~\cite{tseacc25}. 


	\section{Conclusion}
\label{section:conclusion}
This paper proposes \approach, a distilled lifelong planning framework with ranked workload similarity analysis and weighted configuration seeding components for self-adapting configurable systems. The goal is to dynamically determine when it is generally more promising to seed with historical knowledge (\textbf{\textit{when to seed?}}) and extract what knowledge should be redirected for planning without injecting misleading information (\textbf{\textit{what to seed?}}). Empirical studies conducted on nine real-world configurable systems, spanning various domains and encompassing a total of 93 workloads, demonstrate that compared with state-of-the-art approaches, \approach~is:

\begin{itemize}
	\item \textbf{more effective,} as it achieves considerably better adaptation planning than its four competitors with up to 2.29$\times$.
	
	\item \textbf{more efficient,} as it exhibits up to 2.22$\times$ speedup on producing promising configurations.
\end{itemize}

This work sheds light on the importance of automatically leveraging distilled past knowledge to self-adapt configurable systems in planning for future workloads. Looking ahead, we aim to delve into landscape analysis methods to better handle workload evolution and explore feedback mechanisms for more precise identification of beneficial planning information.

\section*{Acknowledgement}
This work was supported by a NSFC Grant (62372084) and a UKRI Grant (10054084).

	\balance
	
\bibliographystyle{IEEEtran}
\bibliography{IEEEabrv,references}

\end{document}